\newcommand{\diracslash}[1]{#1\llap{/\kern2pt}}

\newcommand{\be}{\begin{equation}}
\newcommand{\ee}{\end{equation}}
\newcommand{\bea}{\begin{eqnarray}}
\newcommand{\eea}{\end{eqnarray}}
\newcommand{\ba}[1]{\begin{array}{#1}}
\newcommand{\ea}{\end{array}}

\newcommand{\bt}{\begin{tabular}}
\newcommand{\et}{\end{tabular}}

\newcommand{\beas}{\begin{eqnarray*}}
\newcommand{\eeas}{\end{eqnarray*}}

\documentclass[preprint,prd,aps,floats,nofootinbib,floatfix]{revtex4}
\usepackage{graphicx}
\begin{document}

\title {In-medium decay widths of hidden and open charm vector mesons
in a field theoretic model for composite hadrons}
\author{Amruta Mishra}
\email{amruta@physics.iitd.ac.in; mishra@th.physik.uni-frankfurt.de}
\affiliation{Department of Physics, Indian Institute of Technology, Delhi,
Hauz Khas, New Delhi -- 110 016, India}

\author{S.P. Misra}
\email{misrasibaprasad@gmail.com}
\affiliation{Institute of Physics, Bhubaneswar -- 751005, India} 

\author{W. Greiner
\footnote {email: greiner@fias.uni-frankfurt.de}}
\address{
Frankfurt Institute for Advanced Studies
Ruth-Moufang-Strasse 1,
Frankfurt am Main, D-60438, Germany}

\begin{abstract}
We calculate the decay widths of the charmonium states, $J/\psi$,
$\psi (3686)$ and $\psi(3770)$, to $D\bar D$ pairs, 
as well as the decay width of $D^* \rightarrow D\pi$,
in isospin asymmetric strange hadronic matter, 
using a field theoretical model for composite
hadrons with quark constituents. For this purpose 
we use the quark antiquark pair creation term of
the free Dirac Hamiltonian written in terms of the
constituent quark field operators, and use explicit charmonium, 
$D$, $\bar {D}$, $D^*$ and $\pi$ states to evaluate the matrix 
elements for the charmonium as well as $D^*$ decay amplitudes. 
The medium modifications of the partial decay widths of charmonium
to $D\bar D$ pair, arising
from the mass modifications of the $D(\bar D)$ and the charmonium
states calculated in a chiral effective model, are also included.
The results of the present investigations are then compared with 
the decay widths computed earlier, in a model using light quark 
pair creation in $^3P_0$ state. As in $^3P_0$ model,
the decay amplitude in the present model is multiplied with 
a strength parameter for the light quark pair creation, 
which is fitted from the observed vacuum 
decay width. The effects of the isospin asymmetry, the strangeness fraction
of the hadronic matter on the masses of the charmonium states
and $D(\bar D)$ mesons and hence on the decay widths, have also been
studied. The isospin asymmetry effect is observed
to be dominant for high densities, leading to appreciable difference 
in the decay channels of the charmonium to $D^+ D^-$ and 
$D^0 \bar {D^0}$ pairs. The decay width of $D^* \rightarrow D \pi$
in the hadronic matter has also been calculated within the
composite quark model in the present work, accounting for
the medium modifications of the $D$ and $D^*$ masses. 
The density modifications of the charmonium states and 
$D(D^*)$ mesons, which are observed to be appreciable 
at high densities, will be of relevance in the compressed 
baryonic matter (CBM) experiments at the future facility 
of FAIR, GSI, where charmed hadrons will be produced  
by annihilation of antiprotons on nuclei. 
The interactions of the charmonium states and $D(D^*)$ with the nuclear
medium could lead to the possiblility of the formation of exotic bound states 
of the nuclei with the (excited) charmonium states as well as 
with $D(D^*)$ mesons.

\end{abstract}

\maketitle

\def\bfm#1{\mbox{\boldmath $#1$}}
\def\bfs#1{\mbox{\bf #1}}

\section{Introduction}
The study of the properties of hadrons in the medium is an important 
and challenging topic of research in strong interaction physics. 
The topic is of direct relevance in the context of heavy ion collision 
experiments, which probe matter under extreme conditions, for example,
at high temperatures and/or densities. The properties of the hadrons 
as modified in the medium affect the experimental observables
from the strongly interacting matter produced in heavy ion collision
experiments. The properties of the kaons and antikaons in the medium 
are of relevance in neutron star phenomenology where an attractive 
interaction of antikaon-nucleon can lead to antikaon condensation 
in the interior of the neutron stars \cite{kaplan1,kaplan2}. The medium 
modifications of kaons and antikaons can also affect observables like,
the production as well as collective flow of the kaons and antikaons,
in the heavy ion collision experiments. 
The modifications of the masses of the charm mesons, $D$ and 
$\bar D$ as well as the $J/\psi$ mesons and the excited states of 
charmonium, can have important consequences on the yield of open 
charm mesons as well as of $J/\psi$ in heavy ion collision experiments.
Also, in high energy heavy ion collision experiments at RHIC as well as LHC,
the formation of the quark-gluon-plasma (QGP) \cite{blaiz, satz} 
can lead to the suppression of $J/\psi$.

The $D$ ($\bar D$) mesons are made up of one heavy charm quark 
(antiquark) and one light (u or d) antiquark (quark). 
In the QCD sum rule approach, the modifications of the 
masses of the $D$ ($\bar D$) mesons in the hadronic matter are due
to the interactions of light antiquark (quark) present in the $D$($\bar D$) 
mesons with the light quark condensate \cite{haya1,qcdsum08}. 
The properties of the $D(\bar D)$ meson have also been studied
using the quark meson coupling (QMC) model \cite{qmc1,qmc2} as well as 
the coupled channel approach \cite{ltolos,ljhs,mizutani6,mizutani8,HL}. 
The $D$ and $B$ mesons have been studied within a hadronic framework
by using pion exchange \cite{sudoh}, which leads to attractive
interaction of the $\bar D$ and $B$ mesons in the nuclear matter,
suggesting that these can form bound states with the atomic nuclei.
In a recent study \cite{dnkrein}, the $\bar D$-nucleon interactions  
have been studied using a quark model for the hadrons, in which the
baryons/mesons are constructed as bound states of the constituent quarks
(antiquarks). The field operators of the constituent quarks are 
written in terms of a constituent quark mass, 
which arises from dynamical chiral symmetry breaking 
\cite{dnkrein,hmamspm1,hmamspm2,spmeffpot,spmbothcond}.

Using the QCD sum rule approach, the masses of the charmonium states, 
which are made up of a heavy charm quark and a charm antiquark, 
are due to their interaction with the gluon condensates 
\cite{kimlee,charmmass2}.
Within a chiral effective model, the gluon condensates of QCD,
are simulated by a scalar dilaton field \cite{charmmass2,amcharmdw},
and the in-medium masses of the charmonium states are studied
by the in-medium changes of the dilaton field within the
effective hadronic model.
In the present work, we study the medium modification of the decay widths 
of $J/\psi$ and the excited charmonium states $\psi(3686)$ and $\psi(3770)$ 
to $D\bar D$ pairs, as well as of the decay width of $D^*\rightarrow D\pi$,
in the strange hadronic medium, due to the mass modifications 
of these hadrons calculated in the effective chiral model
\cite{amcharmdw}. These in-medium decay widths are studied
using a field theoretical model for the composite hadrons 
with quark constituents \cite{spm781,spm782,spmdiffscat}.

The outline of the paper is as follows: In section II, we give 
a very brief introduction of the field theoretical model for the 
hadrons with quark constitutents using explicit constructions 
of the charmonium states, the $D^*$ meson as well as the $D$ 
and $\bar D$ mesons in terms of the quark and antiquark constituents.
We then calculate the matrix element of the S-matrix in the lowest order
to compute the decay width of the charmonium states to $D^+ D^-$ or 
$D^0\bar {D^0}$ pairs 
as well as of $D^* \rightarrow D\pi$.
The matrix element, however, is multiplied 
with a parameter, which is fitted from the observed vacuum decay width
of $\psi(3770)\rightarrow D\bar D$ or $D^*\rightarrow D\pi$. 
These decay widths have been calculated
using the field theoretic model for composite hadrons and 
their medium modifications
have been studied as arising from the changes in the masses
of these mesons in the hadronic medium.
In section III, we briefly describe the chiral effective 
model, which has been used to investigate the in-medium masses of 
the open charm mesons ($D (D^0, D^+)$ and $\bar D (\bar {D^0},D^-)$) 
and of the charmonium states \cite{amcharmdw}. 
The in-medium properties of the $D$ 
and $\bar D$ mesons arise due to their interactions with the 
light hadrons, i.e., the nucleons, hyperons and scalar mesons. 
Within the chiral effective model, the scale symmetry breaking of QCD 
has also been incorporated through a scalar dilaton field. 
The masses of the charmonium 
states in the hadronic medium have been calculated from the modification 
of the gluon condensates \cite{amcharmdw} arising within the effective 
hadronic model from the modification of the dilaton field in the strange 
hadronic matter. In section IV, we discuss the results obtained 
in the present investigation and compare with the existing results
of the partial decay widths \cite{amcharmdw,friman} using 
a quark pair creation model, namely the $^3 P_0$ model 
\cite{yopr1,yopr2,yopr3,barnes}.
In the $^3P_0$ model, a light quark antiquark pair is assumed to be 
created in the $^3 P_0$ state, and the light quark (antiquark) 
combines with the heavy charm 
antiquark (charm quark) of the decaying charmonium state to produce
the open charm $D$ and $\bar D$ mesons. It might be noted here that
in the composite hadron model as used in the present work, 
the decay widths of the charmonium state to $D\bar D$ pair, as well as 
of the decay process $D^*\rightarrow D\pi$ in the hadronic medium, 
are calculated using the quark pair creation term of the free Dirac 
Hamiltonian written in terms of the constituent quark field operators 
within the model. 
As has already been mentioned, the matrix element for the 
specific decay process has to be multiplied with a parameter 
corresponding to the quark pair creation, which is fitted 
from the observed vacuum decay width.
In section V, we summarize 
the results for the medium modifications of these decay widths,
calculated in the present field theoretic model for the composite 
hadrons with quark/antiquark constituents, and discuss possible outlook.

\section{The model for composite hadrons}

We shall very briefly discuss the model to clarify the notations, 
so as to apply the same in the present problem.

The field operator at t=0 for a constituent quark for a hadron at rest
is taken as
\begin {equation}
\psi ({\bf x},t=0)=Q({\bf x})+\tilde Q({\bf x}),
\end{equation}
where $Q({\bf x})$ and $\tilde Q({\bf x})$ are the quark annihilation
and antiquark creation operators and are given as
\begin{equation}
Q({\bf x})=(2\pi)^{-{3}/{2}}{\int {U({\bf k}) Q_I ({\bf k})
\exp(i{\bf k} \cdot{\bf x})d\bfs k}}
\label{qx}
\end{equation}
and
\begin{equation}
\tilde Q({\bf x})=(2\pi)^{-{3}/{2}}
{\int {V({\bf k}) \tilde Q_I ({\bf k})
\exp(-i{\bf k} \cdot{\bf x})d\bfs k}}.
\label{tldqx}
\end{equation}
In the above, $Q_I({\bf k})$ and $\tilde Q _I ({\bf k})$ are the two component
quark annihilation and antiquark creation operators, given as
$Q_I({\bf k})=Q_{Ir}({\bf k})u_{Ir}$ and 
$\tilde Q _I ({\bf k})=\tilde Q_{Is}({\bf k})v_{Is}$. The summation over
the dummy indices is understood. $Q_{Ir}({\bf k})$ annihilates a quark
with spin $r$ and momentum ${\bf k}$ and $\tilde Q _{Is}({\bf k})$
creates an antiquark with spin $s$ and momentum ${\bf k}$ and they satisfy
the usual anticommutation relations
\begin{equation}
\{Q_{Ir}({\bf k}),Q_{Is}({\bf k}')^\dagger\}=
\{\tilde Q_{Ir}({\bf k}),\tilde Q_{Is}({\bf k}')^\dagger\}=
\delta _{rs} \delta ({\bf k}-{\bf k}')
\end{equation}
In equations (\ref{qx}) and (\ref{tldqx}), $U(k)$ and $V(k)$ are given as
\begin{equation}
U(k)=\left (\begin{array}{c} f({{\bf k}}^2)\\
{\bfm\sigma}\cdot {\bf k} g({{\bf k}}^2)\\
\end{array} \right ),\;\;\;\;\;
V(k)=\left (\begin{array}{c} 
{\bfm\sigma}\cdot {\bf k} g({{\bf k}}^2)\\
f({{\bf k}}^2)\\
\end{array} \right ),
\label{ukvk}
\end{equation}
where the equal time anticommutation relation for the four-component Dirac field
operators gives the constraint \cite{spm781}
\begin{equation}
f^2+g^2 {\bf k}^2=1,
\label{fgk}
\end{equation}
on the functions $f({\bf k})$ and $g({\bf k})$. For free Dirac field
of mass $M$, we have 
\begin{equation}
f({\bf k})=\left ( \frac{k_0 +M}{2 k_0}\right )^{1/2},\;\;\;\; 
g({\bf k})=\left ( \frac{1}{2 k_0 (k_0+M)}\right )^{1/2},
\label{fkgk}
\end{equation}
where $k_0=(k^2+M^2)^{1/2}$. 
It may be relevant here to note that  when the constituent quark
mass arises from dynamical symmetry breaking, the constituent
quark mass, $M$ in equation (7) in general could be momentum 
dependent \cite{dnkrein,hmamspm1,hmamspm2,spmeffpot,spmbothcond}. 
In a color confining model \cite{dnkrein}, the solution of the gap
equation arising from dynamical chiral symmetry breaking leads
to the constituent quark mass, $M(k)$, which is observed 
to change appreciably only at large momenta \cite{dnkrein}. 
On the other hand, in Nambu Jona Lasinio model
with four point inetraction, such a mass 
is momentum independent and is calculated self-consistently
by solving a gap equation.
In the present work for the study of charmonium and $D^*$ 
decay widths with light quark pair creation, 
we shall assume $M(k)=M$, a constant.
We take the constituent quark mass, $M$ for the 
light quark (u,d) as an input parameter and the value of 
$M_u=M_d$ is chosen to be 330 MeV, a typical value obtained 
from dynamical chiral symmetry breaking, e.g., in Nambu-Jona 
Lasinio model \cite{amhm2004}.
We shall also assume a small momentum expansion for the
the functions $f({\bf k})$ and $g({\bf k})$ of the Dirac 
field operator for the constituent quarks, which yields
\begin{eqnarray}
g({\bf k})=\left ( \frac{1}{2 k_0 (k_0+M)}\right )^{1/2}
\approx \frac {1}{2M},\nonumber \\
f({\bf k})=(1-g^2{\bf k}^2)^{1/2} \approx 1-\frac{1}{2}g^2 {\bf k}^2.
\label{fgapprox}
\end{eqnarray}

The above field operators for the quarks are for the constituents 
of hadrons at rest. To describe these as constituents of hadron 
with finite momentum, we need to suitably Lorentz boost these operators, 
which requires the knowledge of the time dependence of the 
field operators in addition to the space dependence as above. 
This was taken to be given by quarks occupying
fixed energy levels \cite{spm781,spm782} as in the bag model,
so that for the $i$-th quark we have
\begin{equation}
Q_i(x)=Q_i({\bf x})\exp(-i\lambda_i M t),
\label{thadrest}
\end{equation}
where $\lambda_i$ is the fraction of the energy (mass) of the hadron 
carried by the quark, with $\sum_i \lambda_i=1$.
Eq. (\ref{thadrest}) is for hadrons at rest, and, for a hadron in motion
with four momentum p, with appropriate Lorentz boosting \cite{spmdiffscat}
the field operator for quark annihilation for t=0 is given as
\begin{equation}
Q^{(p)}({\bf x},0)=(2\pi)^{-{3}/{2}} {\int {d\bfs k S(L(p)) U({\bf k}) 
Q_I ({\bf k}+\lambda {\bf p})\exp(i({\bf k}+\lambda {\bf p}) 
\cdot{\bf x})}}.
\label{qxp}
\end{equation}
For the antiquark creation operator we similarly have for t=0
\begin{equation}
\tilde Q^{(p)}({\bf x},0)=(2\pi)^{-{3}/{2}} 
{\int {d\bfs k S(L(p)) V(-{\bf k}) 
\tilde Q_I (-{\bf k}+\lambda {\bf p})
\exp(-i(-{\bf k}+\lambda {\bf p}) \cdot{\bf x})}}
\label{tldqxp}
\end{equation}
The form of Lorentz boosting that gives the constituent quark field operators
as above was chosen in Ref  \cite{spmdiffscat} . 

The determination of the $\lambda_i$'s, which correspond to the fractions 
of energy (mass) carried by the constituent quarks in the hadron, will be 
explicitly described in the next section when we construct the $D$-meson states 
with the constituent quarks.

\subsection{Decay widths of the charmonium states to $D\bar D$ pair in the
composite model of the hadrons}

In the present work, we shall study the medium dependence of the partial 
decay widths of the charmonium state, $\psi$ 
($J/\psi$, $\psi(3686)$ and $\psi(3770)$)
to $D\bar D (D^+D^- \; {\rm  or} \; D^0 \bar {D^0}$). 
In the hadronic medium, the masses of the $D$ and $\bar {D}$ are observed
to be different due to the difference in their interactions with the 
hadronic matter. Accounting for this fact, the magnitude of $\bf p$, 
the 3-momentum 
of the $D$ ($\bar D$) meson, when the charmonium state $\psi$ decays 
at rest, is given by
\begin{equation}
|{\bf p}|=\Big (\frac{{m_\psi}^2}{4}-\frac {{m_D}^2+{m_{\bar D}}^2}{2}
+\frac {({m_D}^2-{m_{\bar D}}^2)^2}{4 {m_\psi}^2}\Big)^{1/2}.
\label{pd}
\end{equation}
We note that the masses of $\psi'' \equiv \psi(3770)$ and $D^\pm$ 
in the vacuum are given as
\begin{equation}
m_{\psi''}=3773\;{\rm MeV};\;\;\;\;\; m_{D^\pm}=1869\;{\rm MeV}
\label{mass}
\end{equation}
so that this decay is admissible in vacuum. 
The in-medium effects of the decay widths of the
charmonium state ($\psi \equiv J/\psi, \psi'\equiv \psi(3686), 
\psi''\equiv \psi(3770)$) in the
present work are calculated by considering the medium modifications
of the masses of the charmonium state and $D^\pm$ mesons.

We write the charmonium state $\psi$ ($J/\psi$, $\psi(3686)$ 
and $\psi(3770)$) with spin projection $m$ at rest as
\begin{equation}
|\psi_m(\vec 0)\rangle = {\int {d {\bf k}_1 {c_I} ^i ({\bf k}_1)^\dagger
a_m(\psi,{\bf k}_1)\tilde {c_I}^i (-{\bf k}_1)|vac\rangle}},
\label{psi}
\end{equation}
where, $i$ is the color index of the quark/antiquark operators.
For $\psi\equiv J/\psi$,
\begin{equation}
a_m(\psi,\bfs k_1)\equiv  \bfm \sigma_m u_{J/\psi} (\bfm {k_1})
=\bfm\sigma_m\frac{1}{\sqrt{6}}
\left (\frac {{R_{\psi}}^2}{\pi} \right)^{3/4}
\exp \Big(-\frac {{R_{\psi}}^2 {{\bf k_1}}^2}{2}\Big),
\label{amjpsi}
\end{equation}
for $\psi \equiv \psi'$,
\begin{equation}
a_m(\psi,\bfs k_1)\equiv  \bfm \sigma_m u_{\psi'} (\bfm {k_1})
=\bfm\sigma_m\frac{1}{\sqrt{6}}
{\sqrt {\frac{3}{2}}}
\Big({\frac{R_{\psi'}^2}{\pi}}\Big)^{3/4}
\left(\frac{2}{3}R_{\psi'}^2\bfs k_1^2-1\right)
\exp\left[-\frac{1}{2}R_{\psi'}^2\bfs k_1^2\right].
\label{ampsip}
\end{equation}
and, for $\psi \equiv \psi''$ \cite{spmddbar80},
\begin{equation}
a_m(\psi,{\bf k}_1)=\frac {1}{4\sqrt {3\pi}} u_{\psi''}({\bf k}_1)
\Big ( \bfm\sigma_m -3 (\bfm\sigma \cdot \hat {k_1})\hat {k_1}^m \Big ),
\label{ampsipp}
\end{equation}
where,
\begin{equation}
u_{\psi''}({\bf k}_1)=\Big(\frac{16}{15}\Big)^{1/2} \pi^{-{1}/{4}}
(R_{\psi''}^2)^{7/4} {{\bf k}_1}^2 \exp \Big (-\frac{1}{2} {R_{\psi''}}^2 
{{\bf k}_1}^2\Big).
\label{upsipp}
\end{equation}
In the above, we have taken harmonic oscillator wave functions
for the charmonium states, where $J/\psi$, $\psi'$ and $\psi''$
correspond to the 1S, 2S and 1D states respectively. 
The $D^+$ and $D^-$ states, with finite momenta, 
are explicitly given as
\begin{equation}
|D^+ ({\bf p})\rangle ={ \int {{c_I}^{{i_1}}({\bf k}_2+\lambda_2 {\bf p})
^\dagger u_{D^+}({\bf k}_2)\tilde {d_I}^{i_1} 
(-{\bf k}_2 +\lambda_1 {\bf p}) d\bfs k_2}}
\label{dp}
\end{equation}
and 
\begin{equation}
|D^- ({\bf p}')\rangle = { \int {{d_I}^{{i_2}}({\bf k}_3+\lambda_1 {\bf p}')
^\dagger u_{D^-}({\bf k}_3)\tilde {c_I}^{i_2} 
(-{\bf k}_3 +\lambda_2 {\bf p}') d\bfs k_3}}.
\label{dm}
\end{equation}
In the above, 
\begin{equation}
u_{D^\pm}({\bf k})=\frac{1}{\sqrt{6}}\Big (\frac {{R_D}^2}{\pi} \Big)^{3/4}
\exp\Big(-\frac {{R_D}^2 {{\bf k}}^2}{2}\Big).
\label{udpm}
\end{equation}
For the above states we have used alternative Lorentz boosting,  
which is like getting the hadron through translation operator, 
and is given by equations (\ref{qxp}) and (\ref{tldqxp}) 
\cite{spmdiffscat}.
We shall now explicitly calculate $\lambda_1$ and $\lambda_2$. We recall
that it was conjectured in Ref \cite{spm782} that the binding energy of 
the hadron 
as shared by the quarks shall be {\it inversely} proportional to the quark
masses. Thus we shall explicitly have, for the energies of
$d(\bar d)$ and $\bar c (c)$ in $D(\bar D)$ meson as
\cite{spm782},
\begin{equation}
\omega_1=M_d+\frac{M_c}{M_c+M_d}\cdot(m_D-M_c-M_d)
\label{omega1}
\end{equation}
and,
\begin{equation}
\omega_2=M_c+\frac{M_d}{M_c+M_d}\cdot(m_D-M_c-M_d),
\label{omega2}
\end{equation}
with 
\begin{equation}
\lambda_i=\frac{\omega_i}{m_D}.
\label{lambda12}
\end{equation} 

We next evaluate the matrix element of the quark-antiquark pair creation
part of the Hamiltonian,
between the initial charmonium state and the final state for the reaction
$\psi \rightarrow D^+ ({\bf p})+D^-({\bf p}')$.

The relevant part of the quark pair creation term is through the
$d \bar d$ creation.
From equations (\ref{qxp}) and (\ref{tldqxp})
we can write down
${\cal H}_{d^\dagger\tilde d}({\bf x},t=0)$, and then
integrate over ${\bf x}$ to obtain the expression  
\begin{eqnarray}
&& {\int {{\cal H}_{d^\dagger\tilde d}({\bf x},t=0)d{\bf x}}}
\nonumber\\
&=&
{\int {d\bfs k d\bfs k'd_I^i({\bf k}+\lambda_1{\bf p}')^\dagger 
U({\bf k})^\dagger S(L(p'))^\dagger
\delta(-{\bf k}'+\lambda_1{\bf p}+{\bf k}+\lambda_1{\bf p}')}}
\nonumber\\
&& (\bfm \alpha \cdot ({\bf k}+\lambda_1{\bf p}')+\beta M_d)
S(L(p))V(-{\bf k}')\tilde d_I^i(-{\bf k}'+\lambda_1{\bf p}),
\label{hint}
\end{eqnarray}
where $M_d$ is the constituent mass of the $d$ quark.
In equation (\ref{hint}), the Lorentz boosting factor $S(L(p))$
is given as 
\begin{equation}
S(L(p))=\left ( \frac{p^0+m_h}{2m_h}\right )^{1/2}
+\frac {\bfm\alpha \cdot {\bf p}}{(2m_h (p^0+m_h))^{1/2}},
\label{slp}
\end{equation}
with $m_h$ is the mass of the hadron with momentum ${\bf p}$,
which is the $D$ meson here.

From equation (\ref{hint}) we can then evaluate that
\begin{equation}
\langle D^+ ({\bf p}) | \langle D^- ({\bf p}')|
{ \int {{\cal H}_{d^\dagger\tilde d}({\bf x},t=0)d{\bf x}}}
|{\psi }_m (\vec 0) \rangle
=\delta({\bf p}+{\bf p}'){\int {d\bfs k_1 
A^{\psi}_m({\bf p},{\bf k}_1)}},
\label{psidd}
\end{equation}
with appropriate simplifications using equations for
the states $|\psi_m({\bf 0})\rangle$, $|D^+({\bf p})\rangle$
and $|D^-({\bf p'})\rangle$ given in (\ref{psi}),
(\ref{dp}) and (\ref{dm}). In the above equations the
spectators $c$ and $\tilde c$ give that
\begin{equation}
{\bf k}_2+\lambda_2{\bf p}={\bf k}_1; \;\;\; 
-{\bf k}_3+\lambda_2{\bf p}'=-{\bf k}_1.
\label{cc}
\end{equation}
Also, $d$ and $\tilde d$ contractions of the above with the same 
in (\ref{hint}) yield the results
\begin{equation}
{\bf k}+\lambda_1 {\bf p}'={\bf k}_3+\lambda_1{\bf p}';\;\;\; 
-{\bf k}'+\lambda_1{\bf p}=-{\bf k}_2+\lambda_1{\bf p}.
\label{dbard}
\end{equation}
This enables us to integrate over all momenta except ${\bf k}_1$.
For the evaluation of 
$A^{\psi}_m({\bf p},{\bf k}_1)$, we first note that during the 
evaluation of the above matrix element, we have a spin matrix 
given as $a_m(\psi,{\bf k}_1)$ as well as a spin matrix in 
the expression for the pair creation term in equation 
(\ref{hint}). For the evaluation of $A^{\psi}_m({\bf p},{\bf k}_1)$ 
we get, including summing over color,
\begin{eqnarray}
 A^{\psi}_m({\bf p},{\bf k}_1)
&=& 3 u_{D^+}({\bf k}) u_{D^-}({\bf k})
\cdot {\rm {Tr}} \big [a_m(\psi,{\bf k}_1)
U({\bf k})^\dagger  \nonumber\\
&& S(L(p'))^\dagger
(\bfm\alpha\cdot({\bf {\tilde q}})+\beta M_d)
S(L(p))V(-{\bf k})
\big ],
\label{ampk}
\end{eqnarray}
where, ${\bfm k} = {\bf k}_1 -\lambda_2{\bf p}$,
${\bfm {\tilde q}} = {\bf k}_1 -{\bf p}$ and
$\bfm p'=-\bfm p$.
We shall now simplify $A^{\psi}_m({\bf p},{\bf k}_1)$. Firstly, since the 
$D(\bar D)$ mesons are completely nonrelativistic, we shall be taking 
that $S(L(p))$ and $S(L(p'))$ are identity. $U,V$ are given as in equation 
(\ref{ukvk}). The integral in the R.H.S. of the equation (\ref{psidd})
can be written as,
\begin{eqnarray}
\int d {\bfm {k_1}} A^{\psi}_m({\bf p},{\bf k}_1)
=3\int d {\bfm {k_1}} u_{D^+}({\bf k}) u_{D^-}({\bf k})
\cdot {\rm {Tr}} \big [a_m(\psi,{\bf k}_1)
B({\bfm k}, {\bfm {\tilde q}})
\big ],
\label{ampk1}
\end{eqnarray}
where,
\begin{equation}
B(\bfm k, {\bfm {\tilde q}})={\bfm \sigma}\cdot {\bfm {\tilde q}}
-\Big ( 2 (\bfm k \cdot {\bfm {\tilde q}})g^2 +f(\bfm k) \Big )
{\bfm \sigma}\cdot {\bfm k}.
\end{equation}
We use the approximate forms of $f$ and $g$, 
$f\approx 1-\frac {g^2 {\bf k}^2}{2}$, and $2M_d g \approx 1$,
of the equation (\ref{fgapprox}) for simplifying the integral
(\ref{ampk1}). The integral (\ref{ampk1}) can be simplified to
\begin{eqnarray}
\int d {\bfm {k_1}} A^{\psi}_m({\bf p},{\bf k}_1)
=6c_\psi \exp\big [(a_\psi {b_\psi}^2 -{R_D}^2 
{\lambda_2}^2){\bfm p}^2\big]
\int d {\bfm {k_1}} T^{\psi}_m({\bf p},{\bf k}_1),
\end{eqnarray}
where, 
$T^{\psi}_m({\bf p},{\bf k}_1)$ is given as,
\begin {equation}
T^{\psi}_m({\bf p},{\bf k}_1) 
=\frac{1}{2} {\rm {Tr}} \big [\sigma_m
B({\bfm k}, {\bfm {\tilde q}})\big],
\end{equation}
for $\psi=J/\psi$,
\begin {equation}
T^{\psi}_m({\bf p},{\bf k}_1) 
=\frac{1}{2} {\rm {Tr}} \big [\sigma_m
B({\bfm k}, {\bfm {\tilde q}})\big]
\cdot \left(\frac{2}{3}R_{\psi'}^2\bfs k_1^2-1\right),
\end{equation}
for $\psi=\psi'$, and,
\begin {equation}
T^{\psi}_m({\bf p},{\bf k}_1) 
=\frac{1}{2} {\bfm {k_1}}^2 {\rm {Tr}} 
\big [\left(\bfm\sigma_m-3\hat k_1^m(\bfm\sigma\cdot\hat k_1)
\right)\cdot
B({\bfm k}, {\bfm {\tilde q}})\big],
\end{equation}
for $\psi=\psi''$. In the above, the parameters $a_\psi$, $b_\psi$
and $c_\psi$ are given as
\begin{equation}
a_\psi=\frac{1}{2}R_{\psi}^2+R_D^2; \;\;\;\; 
b_\psi=R_D^2\lambda_2/a_\psi,
\label{abpsi}
\end{equation}
with $R_\psi$ as the radius of the charmonium state,
$\psi =J/\psi,\psi',\psi''$, and,
\begin{equation}
c_{J/\psi}=\frac{1}{\sqrt{6}}\cdot
\left(\frac{R_\psi^2}{\pi}\right)^{\frac{3}{4}}
\cdot\frac{1}{6}\cdot\left(\frac{R_D^2}{\pi}\right)
^{\frac{3}{2}},
\label{cpsi}
\end{equation}
\begin{equation}
c_{\psi'}=\frac{1}{\sqrt{6}}\left(\frac{3}{2}\right)^{\frac{1}{2}}
\left(\frac{R_{\psi'}^2}{\pi}\right)^{\frac{3}{4}}
\cdot\frac{1}{6}\cdot\left(\frac{R_D^2}{\pi}\right)^{\frac{3}{2}},
\label{cpsip}
\end{equation}
\begin{equation}
c_{\psi''}=\frac{1}{4\sqrt{3\pi}}
\left ({\frac{16}{15}}\right)^{\frac{1}{2}}
\cdot \pi^{-{\frac{1}{4}}}
\cdot (R_{\psi''}^2)^
\frac{7}{4}\cdot\frac{1}{6}\cdot
\left(\frac{R_D^2}{\pi}\right)^{\frac{3}{2}}.
\label{cpsipp}
\end{equation}
We now change the integration variable to $\bfs q$ 
in equation (\ref{ampk1}) with the
substitution $\bfs k_1=\bfs q+{b_\psi}\bfs p$ and write
\begin{equation}
{\int {A^{\psi}_m({\bf p},{\bf k}_1)d\bfs k_1}}
=6c_\psi\exp[(a_\psi {b_\psi}^2-\lambda_2^2 R_D^2)
{\bf p}^2]\cdot 
{\int {\exp(-{a_\psi}\bfs q^2)T_m^{\psi}d\bfs q}}.
\label{ampqt}
\end{equation}
We now note that the terms which are odd in $\bfs q$ in equation 
(\ref{ampqt}) will vanish.
Also, from rotational invariance we shall have 
${\bf q}_m({\bf q}\cdot{\bf p})\equiv \frac{1}{3}{\bf q}^2{\bf p}_m$, and
$(\bfs q\cdot\bfs p)^2\equiv \frac{1}{3}\bfs q^2\cdot\bfs p^2$. 
This yields that the trace in the integral in equation 
(\ref{ampqt}) to be given in the form
\begin{equation}
T_m^{\psi}(\bfs p,\bfs q)\equiv\big [F_0^\psi(|\bfs p|)
+F_1^\psi(|\bfs p|)\bfs q^2+
F_2^\psi(|\bfs p|)(\bfs q^2)^2\big]\bfs p_m,
\label{tmc}
\end{equation}
In equation (\ref{ap}), the coefficients, $F_i^\psi (i=0,1,2)$
for $\psi\equiv J/\psi,\psi',\psi''$, are given as
\begin{eqnarray}
&&F^{J/\psi}_0=(\lambda_2-1)-2g^2\bfs p^2(b_{J/\psi}-\lambda_2)
\left(\frac{3}{4}b_{J/\psi}^2-(1+\frac{1}{2}\lambda_2)b_{J/\psi}
+\lambda_2-\frac{1}{4}\lambda_2^2\right),
\nonumber\\
&&F^{J/\psi}_1=g^2\left[-\frac{5}{2}b_{J/\psi}+\frac{2}{3}
+\frac{11}{6}\lambda_2\right],
\nonumber\\
&&F^{J/\psi}_2=0,
\label{c012psi}
\end{eqnarray}
\begin{eqnarray}
F^{\psi'}_0&=&\left(\frac{2}{3}R_{\psi'}^2{b_{\psi'}}^2\bfs p^2-1\right)
F^{J/\psi}_0,
\nonumber\\
F^{\psi'}_1&=&\frac{2}{3}R_{\psi'}^2 F^{J/\psi} _0
+\left(\frac{2}{3}R_{\psi'}^2 b_{\psi'}^2\bfs p^2
-1\right)
F^{J/\psi}_1
\nonumber \\ &-&\frac{8}{9}R_{\psi'}^2{b_{\psi'}}g^2\bfs 
p^2\left[\frac{9}{4}b_{\psi'}^2-b_{\psi'}\left(2+\frac{5}{2}\lambda_2
\right)+2\lambda_2+\frac{1}{4}\lambda_2^2\right],\nonumber\\
F^{\psi'}_2&=&\frac{2}{3}R_{\psi'}^2g^2
\left[-\frac{7}{2}b_{\psi'}+\frac{2}{3}+\frac{11}{6}
\lambda_2\right],
\label{c012psip}
\end{eqnarray}
and,
\begin{eqnarray}
F_0^{\psi''}&=&2b_{\psi''}^2(1-\lambda_2)\bfs p^2\nonumber \\
&+&2b_{\psi''}^2g^2(\bfs p^2)^2(b_{\psi''}-\lambda_2)((3/2)b_{\psi''}^2
-(2+\lambda_2)b_{\psi''}+2\lambda_2-(1/2)\lambda_2^2),\nonumber\\
F_1^{\psi''}&=&g^2\bfs p^2[14{b_{\psi''}}^3-b_{\psi''}^2((32/3)
+(37/3)\lambda_2)\nonumber \\
&+&b_{\psi''}((28/3)\lambda_2-(1/3)\lambda_2^2)],
\nonumber\\
F_2^{\psi''}&=&g^2[7b_{\psi''}-(2/3)\lambda_2-(4/3)].
\label{c012psipp}
\end{eqnarray}
The integration over $\bfs q$ in equation (\ref{ampqt}) is 
straightforward. On performing the integration, one obtains that
\begin{equation}
{\int {A^{\psi}_m({\bf p},{\bf k}_1)d\bfs k_1}}
 =A^{\psi}(|{\bf p}|){\bf p}_m,
\label{ap3}
\end{equation}
where, $A^{\psi}(|\bfs p|)$ is given as
\begin{equation}
A^{\psi}(|{\bf p}|)=6c_\psi\exp[(a_\psi {b_\psi}^2
-R_D^2\lambda_2^2){\bf p}^2]
\cdot\Big(\frac{\pi}{a_\psi}\Big)^\frac{3}{2}
\Big[F_0^\psi+F_1^\psi\frac{3}{2a_\psi}
+F_2^\psi\frac{15}{4a_\psi^2}\Big].
\label{ap}
\end{equation}
With $<f|S|i>=\delta_4(P_f-P_i)M_{fi}$ we then have for $\psi$ of spin $m$,
\begin{equation}
M_{fi}=2\pi\cdot(-iA^{\psi}(|{\bf p}|)){\bf p}_m.
\label{mfi}
\end{equation}
We shall be investigating the medium effects of the decay width of the
charmonium state to the $D^+D^-$ pair. In the medium, the modifications
of the masses of the outgoing states $D^+$ and $D^-$ are different
because of the difference in their interactions with the hadronic medium.
For the charmonium state decaying at rest, taking the average for spin, 
we obtain the expression for the decay width as
\begin{eqnarray}
\Gamma(\psi\rightarrow D^+D^-)=&& \gamma_\psi^2 \frac{1}{2\pi} 
{\int {\delta(m_{\psi}-p^0_{D^+}-p^0_{D^-})
{|M_{fi}|^2}_{av}
\cdot 4\pi |{\bfs p}_{D^+}|^2 d|{\bfs p}_{D^+}| }}
\nonumber\\
=&& \gamma_\psi^2\frac{8\pi^2}{3}{\bf p}|^3
\frac {{p^0}_{D^+} {p^0}_{D^-}}{m_{\psi}}
A^{\psi}(|{\bf p}|)^2
\label{gammapsidpdm}
\end{eqnarray}
In the above, $p^0_{D^\pm}=\big(m_{D^\pm}^2+{\bf p}^2\big)^{\frac{1}{2}}$, 
and, $|\bfs p|$ is the magnitude of the momentum of the outgoing $D^\pm$ 
mesons. The decay of $\psi$ to $D^0 \bar {D^0}$ proceeds through a 
$u \bar u$ pair creation and the decay width (\ref{gammapsidpdm}) is
modified to 
\begin{eqnarray}
\Gamma(\psi\rightarrow D^0\bar D^0)
=&& \gamma_\psi^2\frac{8\pi^2}{3}\cdot|{\bf p}|^3
\frac {{p^0}_{D^0} {p^0}_{\bar {D^0}}}{m_{\psi}}
A^{\psi}(|{\bf p}|)^2
\label{gammapsid0d0b}
\end{eqnarray}
In the above, $p^0_{D^0}=\big(m_{D^0}^2+{\bf p}^2\big)^{\frac{1}{2}}$, 
$p^0_{\bar {\rm D}^0}=\big(m_{\bar {\rm D}^0}^2+{\bf p}^2\big)
^{\frac{1}{2}}$, 
and, $|\bfs p|$ is the magnitude of the momentum of the outgoing 
$D^0(\bar {\rm D}^0)$ mesons.
In the expressions for the decay widths of the charmonium state,
$\psi$ decaying to $D^+D^- (D^0{\bar {D^0}})$, we have introduced 
the parameter, $\gamma_\psi$, which is the production strength 
of $D\bar D$ from decay of charmonium $\psi$ through light quark 
pair creation. To study the
decay of quarkonia using a light quark pair creation model, namely,
$^3P_0$ model, such a pair creation strength parameter, 
$\gamma$ has been introduced in Ref. \cite{barnes,friman}, 
which was fitted to the observed decay width of the meson.
The parameter, $\gamma_\psi$ in the present work,
is chosen so that one reproduces the vacuum decay widths for 
the decay channels $\psi'' \rightarrow D^+ D^-$ and 
$\psi'' \rightarrow D^0 \bar {D^0}$ \cite{amcharmdw,friman}.
We note that 
$m_{J/\psi}=3097\;{\rm MeV}, m_{\psi'}=3686 \; {\rm MeV}$,
so that the decay of $J/\psi$ or $\psi'$ to $D\bar D$ pair
is not admissible in vacuum. However these decays 
may become kinematically admissible with in-medium effects. 
As we see, the decay widths of the charmonium states are given as
a polynomial part multiplied by a gaussian part.  Hence the medium
dependence of the decay width is due to the combined effect
of the polynomial and the exponential part of the decay width. 

\subsection{Decay width of $D^*\rightarrow D\pi$ in the
composite model of the hadrons}

The masses of $D^*$ and pions are given as 
$m_{D^{*+}}=2010\;{\rm MeV},m_{\pi^+}=139\;{\rm MeV},
m_{\pi^0}=134\; {\rm MeV}$,
so that the decays $D^{*+} \rightarrow D^+\pi^0 (D^0\pi^+)$ 
are possible in vacuum.
We note that for the decay of the $D^*$ meson at rest to $D\pi$, the center 
of mass momentum is given by
\begin{equation}
|\bfs p|=\Bigg (\frac{1}{4}m_{D^{*}}^2-\frac{m_{D}^2+m_{\pi}^2}{2}
+\frac{\left(m_{D}^2-m_{\pi}^2\right)^2}{4m_{D^{*}}^2}
\Bigg )^{1/2}.
\label{modp}
\end{equation}

The appropriate $D^{*+}$ and pion states are given as,
with $q=(u,d)$ doublet,
\begin{eqnarray}
&&|D^{*+}_m(\bfs 0)>=\int d\bfs k_1 c_I^i(\bfs k_1)a_m(D^*,\bfs k_1)
\tilde d_I^i(-\bfs k_1)|vac>,\nonumber\\
&&|\pi^0(\bfs p')>=\int d\bfs k_3 q_I^i(\bfs k_3
+\frac{1}{2}\bfs p')u_\pi(\bfs k_3)
\tau_3\tilde q_I^i(-\bfs k_3+\frac{1}{2}\bfs p')|vac>,
\label{dstarpi}
\end{eqnarray}
where
\begin{eqnarray}
&& a_m(D^*,\bfs k_1)=\frac{1}{\sqrt 6}
\left(\frac{R_{D^*}^2}{\pi}\right)^{3/4}\bfm\sigma_m
\exp\left[-\frac{1}{2}R_{D^*}^2 \bfs k_1^2\right];
\nonumber \\
&& u_\pi(\bfs k_3)=\frac{1}{2\sqrt 3}
\left(\frac{R_{\pi}^2}{\pi}\right)^{3/4}
\exp\left[-\frac{1}{2}R_\pi^2\bfs k_3^2\right].
\label{udstarpi}
\end{eqnarray}
We then obtain the decay width for the above through the pair creation 
term as earlier, given by
\begin{equation}
\langle D^+ ({\bf p}) | \langle \pi^0 ({\bf p}')|
\int{\cal H}_{d^\dagger\tilde d}({\bf x},t=0)d{\bf x}
|D^{*+}_m (\vec 0) \rangle
=\delta({\bf p}+{\bf p}')\int d\bfs k_1 A^{D^{*+}}_m({\bf p},{\bf k}_1),
\label{dstardpi}
\end{equation}
where with the substitution $\bfs k_3=\bfs k_1+\frac{1}{2}\bfs p'$ and
$\bfs k_2=\bfs k_1-\lambda_2 \bfs p$ and momentum conservations as earlier, 
we have,
\begin{eqnarray}
A^{D^{*+}}_m({\bf p},{\bf k}_1)=&& 3\cdot Tr\Big[a_m(D^{*+},{\bf k}_1)u_{\pi^0}(\bfs k_3)
U(\bfs k_3)^\dagger S(L(p'))^\dagger
 (\bfm\alpha\cdot({\bf k}_1-{\bf p})+\beta m_d)\nonumber\\&&
S(L(p))V(-\bfs k_2)u_{D^+}(\bfs k_2)\Big].
\label{amdstrpq}
\end{eqnarray}
As earlier, we take $S(L(p))$ and $S(L(p'))$ as identity and, 
evaluate the integral of $A^{D^{*+}}_m(\bfm p,
\bfm {k_1})$ given by the above equation in the similar way
as was done for the decay of charmonium states to $D\bar D$ pair.
Defining the parameters
\begin{eqnarray}
&&a_{D^*}=\frac{1}{2}\left(R_{D^*}^2+R_D^2+R_\pi^2\right); \;\;\;\; 
b_{D^*}=\frac{1}{2}\left(R_D^2\lambda_2+\frac{1}{2}R_\pi^2\right)/a_{D^*}
\nonumber \\
&&c_{D^*}=\frac{1}{6}\cdot \frac{1}{2\sqrt 3}
\left(\frac{R_{D^*}^2 R_D^2 R_\pi^2}{\pi^3}\right)^{\frac{3}{4}},
\label{abcdstr}
\end{eqnarray}
and, substituting $\bfs k_1=\bfs q+b_{D^*}\bfs p$, we obtain 
\begin{eqnarray}
\int A^{D^*}_m({\bf p},{\bf k}_1)d\bfs k_1
&=&6c_{D^*}\exp\left[a_{D^*}b_{D^*}^2{\bf p}^2
-\frac{1}{2}\left(\lambda_2^2 R_D^2
+\frac{1}{4}R_\pi^2\right){\bf p}^2\right]\nonumber \\ &
\times &\int \exp(-a_{D^*}\bfs q^2)T^{D^*}_md\bfs q.
\label{ampqtdstr}
\end{eqnarray}
Using rotational invariance, the trace $T_m^{D^*}$ is obtained as
\begin{equation}
T^{D^*}_m(\bfs p,\bfs q)\equiv\left[F^{D^*}_0
(|\bfs p|)+F^{D^*}_1(|\bfs p|)\bfs q^2
\right]\bfs p_m,
\label{tmcdstr}
\end{equation}
where,
\begin{eqnarray}
F^{D^*}_0&=&(b_{D^*}-1)\left(1-\frac{1}{2}g^2\bfs p^2
(\lambda_2-\frac{1}{2})^2\right) \nonumber \\
&-&(b_{D^*}-\lambda_2)\left(\frac{1}{2}+g^2\bfs p^2
\left(\frac{3}{4}b_{D^*}^2-\frac{5}{4}b_{D^*}
+\frac{7}{16}\right)\right)
\nonumber\\
&-&(b_{D^*}-\frac{1}{2})\left[\frac{1}{2}+g^2\bfs p^2
\left(\frac{3}{4}b_{D^*}^2-(1+\frac{1}{2}\lambda_2)b_{D^*}
+\lambda_2-\frac{1}{4}\lambda_2^2\right)\right]\nonumber\\
F^{D^*}_1&=&-g^2\left[\frac{5}{2}b_{D^*}-\frac{9}{8}
-\frac{11}{12}\lambda_2\right]
\label{c01dstr}
\end{eqnarray}
On $\bfs k_1$ integration, which results from $\bfs q$ integration, 
we then obtain that
\begin{equation}
\int A^{D^*}_m({\bf p},{\bf k}_1)d\bfs k_1 =A^{D^*}(|{\bf p}|){\bf p}_m
\label{ap3dstr}
\end{equation}
where, 
\begin{eqnarray}
A^{D^*}(|{\bf p}|)&=&6c_{D^*}\exp\left[a_{D^*}b_{D^*}^2{\bf p}^2
-\frac{1}{2}\left(\lambda_2^2 R_D^2
+\frac{1}{4}R_\pi^2\right){\bf p}^2\right] \nonumber \\
&\cdot&\Big(\frac{\pi}{a_{D^*}}\Big)^{3/2}
\Big[F^{D^*}_0+F^{D^*}_1 \Big (\frac{3}{2a_{D^*}}\Big )\Big].
\label{apdstr}
\end{eqnarray}
Now, with $M_{fi}=2\pi i A^{D^*}\bfs p_m$, taking the average 
over the initial spin components, we then obtain that
\begin{eqnarray}
\Gamma\left(D^{*+}\rightarrow D^+\pi^0\right)
&&= \gamma_{D^*}^2\frac{1}{2\pi}
\int\delta(m_{D^{*+}}-p_D^0-p_\pi^0)|M_{fi}|_{av}^2d\bfs p
\nonumber\\
&=& \gamma_{D^*}^2\frac{8\pi^2p_D^0p_\pi^0}{3m_{D^*}}|A^{D^*}
(|\bfs p|)|^2|\bfs p|^3.
\label{gammadstr}
\end{eqnarray}
In the above, $\gamma_{D^*}$ is the production strength
of $D\pi$ from decay of $D^*$ meson, which is fitted from
its vaccum decay width. The medium modification of the
decay width of $D^* \rightarrow D\pi$ is studied as arising 
from the mass modifications of the $D$ and $D^*$ mesons.
From the quark meson coupling model \cite{krein1},
the mass modifications of the $D$ and $D^*$ mesons
have been shown to be nearly the same, and, in the present
investigation, we shall assume $m^*_{D^*}/m^{vac}_{D^*}
=m^*_{D}/m^{vac}_{D}$ to obtain the in-medium changes
of the $D^*$ meson mass in the hadronic medium.

\section{Charmonium states and $D$($\bar D$) mesons in the medium}

We study the medium dependence of the partial decay widths of the 
charmonium  decaying to $D\bar D$ due to medium modifications of the
masses of the $D$ meson, the $\bar D$ meson and the charmonium state
calculated in a chiral effective model \cite{paper3}. The model is based 
on the nonlinear realization of chiral symmetry 
\cite{weinberg,coleman1,coleman2,bardeen} 
and broken scale invariance \cite{paper3,hartree,kristof1}. 
The properties of the hadrons in the strange hadronic matter 
within the chiral effective model are investigated using the 
mean field approximation, where all the meson fields are treated 
as classical fields. By solving the equations of motion of these fields,
the values of the meson fields are obtained. 
These are calculated for given values of the isospin asymmetry parameter,
$\eta= -\frac{\sum_{i}  {I_{3i} \rho_{i}}}{\rho_{B}}$, 
and the strangeness fraction, $f_s= 
\frac{\sum_{i} {s_{i} \rho_{i}}}{\rho_{B}}$, 
where $\rho_i$ is the number density of the baryon of $i$-th type 
($i=p,n,\Lambda,\Sigma^+,\Sigma^0.\Sigma^-,\Xi^-,\Xi^0$) 
and $s_{i}$ is the number of strange quarks in the $i$-th baryon.
Within the chiral effective model, 
the in-medium masses of the $D$ and $\bar D$ mesons \cite{amdmeson} 
have been studied in isospin asymmetric nuclear matter at zero
\cite{amarind} and finite temperatures \cite{amarvind} as well as 
in hot asymmetric hyperonic matter \cite{amcharmdw}.
These arise from the interaction of these mesons with nucleons, 
hyperons and the scalar mesons.  

Within the chiral effective model \cite{amarvind,amcharmdw},
the mass shifts of the charmonium states arise due to interaction 
with the gluon condensates in QCD, simulated 
by a scalar dilaton field through a scale symmetry breaking term
\cite{paper3,amarvind,sche1}.
Equating the trace of the energy momentum tensor in QCD to the
trace of the energy momentum tensor in the chiral effective model
corresponding to the scale symmetry breaking gives
the relation between the gluon condensate to the
dilaton field \cite{charmmass2,cohen,amarvind,amcharmdw}.
In Ref. \cite{charmmass2}, the medium modification of the scalar 
as well as twist 2 gluon condensates 
have been calculated within a chiral effective model 
from the medium modification of a scalar dilaton field, 
which were used to calculate the mass shifts of
the charmonium states, $J/\psi$ and $\eta_c$ 
using the framework of the QCD sum rules \cite{kimlee,charmmass2}. 
The QCD sum rule approach \cite{charmmass2,klingl} and leading order 
perturbative calculations \cite{pes1} have been used to study the 
medium modifications of the masses of the charmonium states, 
from the in-medium values of the gluon condensates calculated 
within the chiral effective model \cite{amcharmdw}, as well as 
from the linear density approximation \cite{leeko}.  
The mass shifts of the charmonium states obtained,
using the chiral effective model \cite{amcharmdw}, at low densities 
are observed to be similar to the calculations of linear density 
approximation \cite{leeko}. 
In the following section, we shall investigate the medium effects of the
charmonium decay widths arising from the in-medium masses
of the $D$, $\bar D$ and charmonium states as calculated
in Ref. \cite{amcharmdw}.

\begin{figure}
\includegraphics[width=12cm,height=12cm]{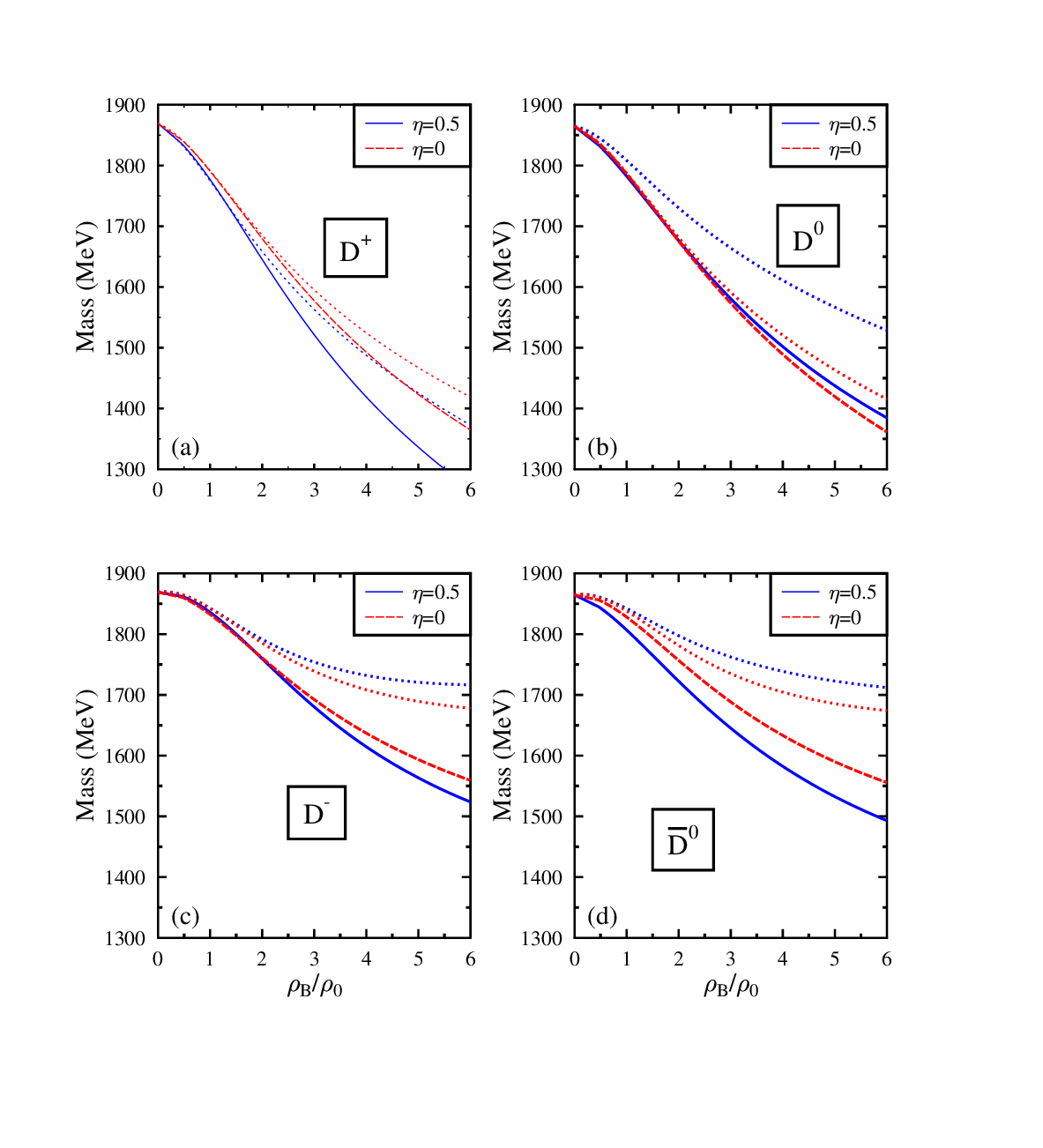}
\caption{(Color online)
The masses of $D$ ($D^+$,$D^0$) and $\bar D$ ($D^-$,$\bar {D^0}$) mesons
are plotted as functions of baryon density in units of nuclear matter
saturation density. These are plotted for the values of the isospin asymmetric 
parameter $\eta=0$ and $\eta$=0.5 for the hyperonic matter with strangeness
fraction,$f_s$=0.5, and are compared with the masses
for the case of nuclear matter ($f_s$=0) shown as dotted lines.
}
\label{mdasym}
\end{figure}

\begin{figure}
\includegraphics[width=10cm,height=10cm]{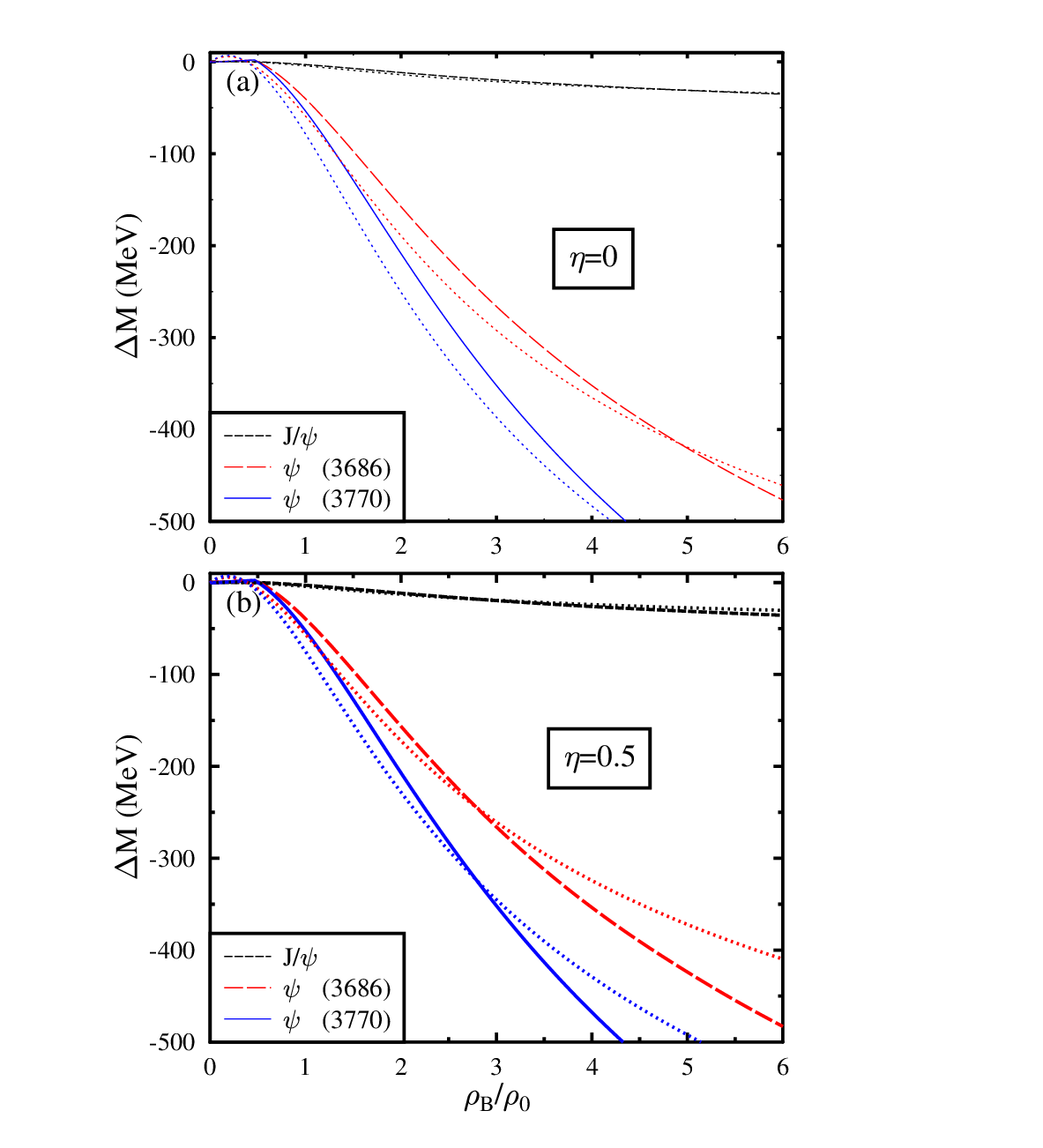}
\caption{(Color online)
The shifts in the masses of the charmonium states ($J/\psi$, $\psi(3686)$
and $\psi(3770)$) are plotted as functions of the baryon density in units
of nuclear matter saturation density. 
These are plotted for the values of the isospin asymmetric 
parameter $\eta=0$ and $\eta$=0.5 and are compared with the masses
for the case of strangeness fraction, $f_s$=0.
}
\label{dmcharmasym}
\end{figure}

\begin{figure}
\vskip -0.5in
\resizebox{0.8\textwidth}{!}{%
  \includegraphics{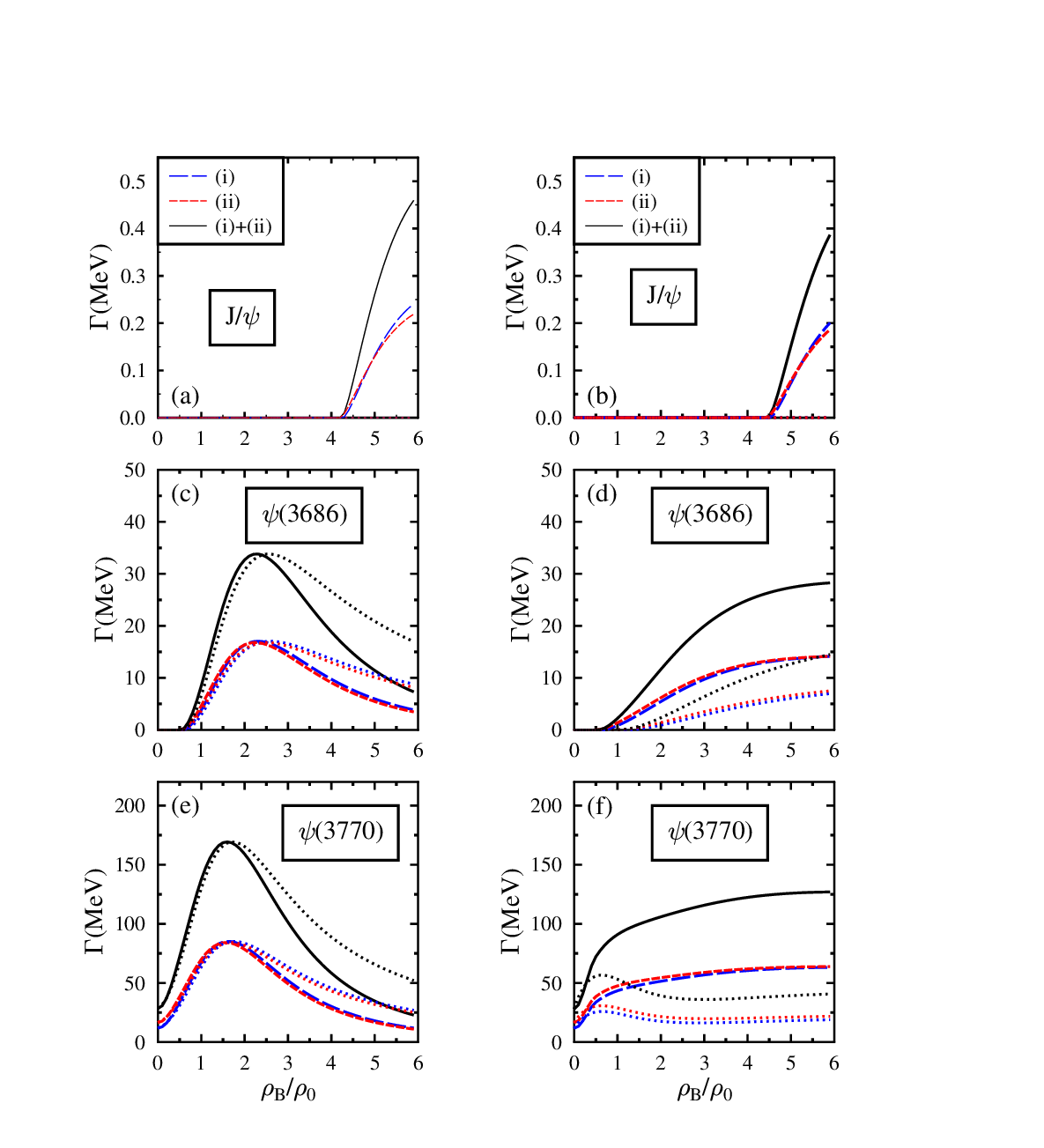}
} 
\caption{(Color online) The partial decay widths of the charmonium states, 
calculated using the present model for composite hadrons,
to (i) $D^+D^-$, (ii) $D^0\bar {\rm D^0}$ and (iii) the sum of the 
two channels ((i)+(ii)) in the isospin symmetric strange hadronic matter 
($\eta$=0,$f_s$=0.5), 
accounting for the medium modifications of the $D(\bar D)$ mesons. 
These are shown in subplots (a), (c) and (e), when the mass modifications 
of the charmonium states are neglected and (b), (d) and (f), 
the partial decay widths are shown when the mass modifications 
of the charmonium states are also taken into account. These results
are compared to the case of nuclear matter ($f_s$=0), shown as dotted lines.}
\label{charmdecayeta0}
\end{figure}

\begin{figure}
\vskip -0.5in
\resizebox{0.8\textwidth}{!}{%
  \includegraphics{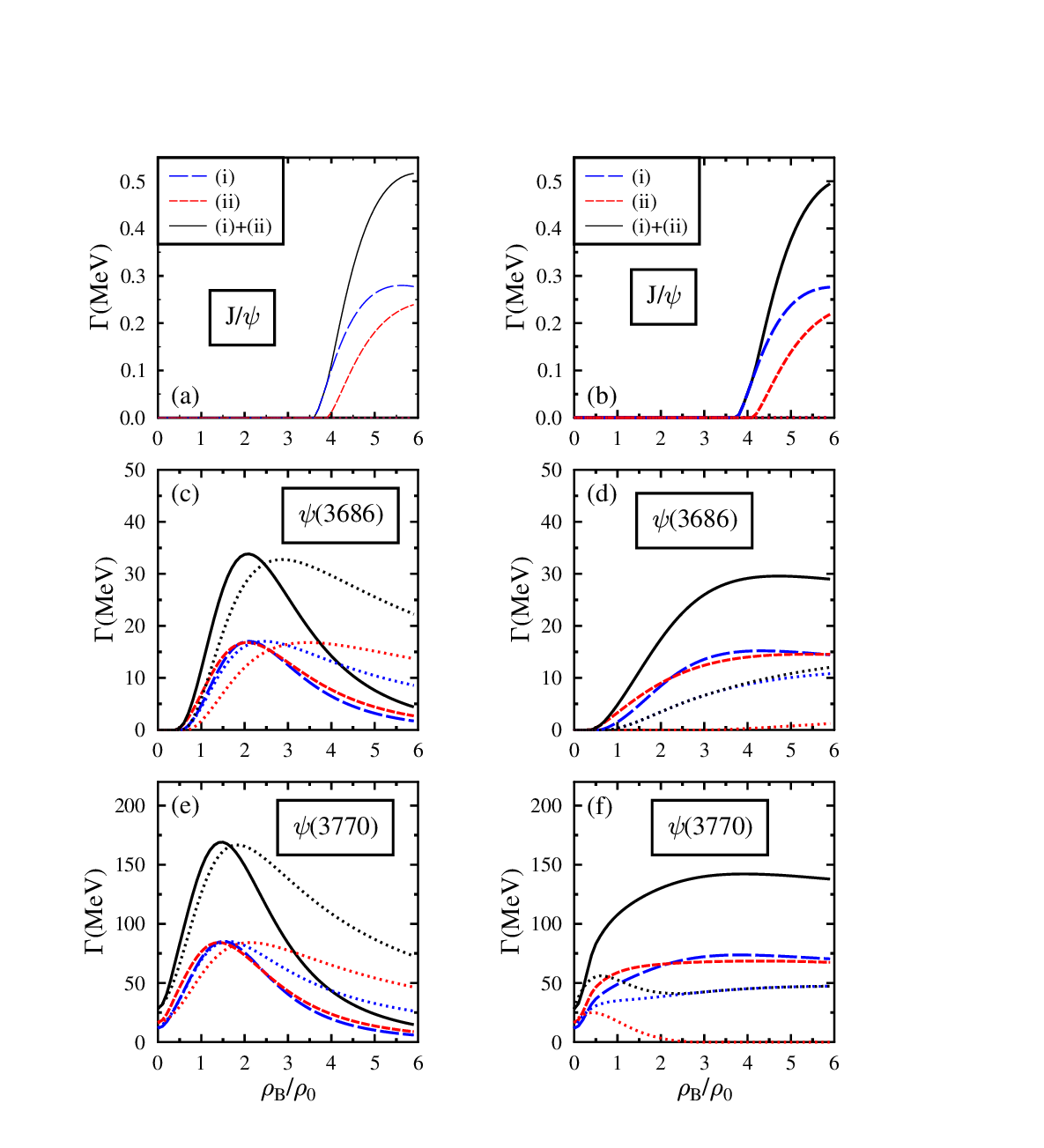}
} 
\caption{(Color online) The partial decay widths of the charmonium states, 
calculated using the present model for composite hadrons,
to (i) $D^+D^-$, (ii) $D^0\bar {\rm D^0}$ and (iii) the sum of the 
two channels ((i)+(ii)) in the isospin asymmetric strange hadronic matter 
($\eta$=0.5, $f_s$=0.5), as functions of the baryon density,
in units of the nuclear matter saturation density, 
accounting for the medium modifications of the $D(\bar D)$ mesons. 
These are shown in subplots (a), (c) and (e), when the mass modifications 
of the charmonium states are neglected and (b), (d) and (f), 
the partial decay widths are shown when the mass modifications 
of the charmonium states are also taken into account. These results
are compared to the case of nuclear matter ($f_s$=0), shown as dotted lines.}
\label{charmdecayeta5}
\end{figure}

\begin{figure}
\vskip -0.5in
\resizebox{0.8\textwidth}{!}{%
  \includegraphics{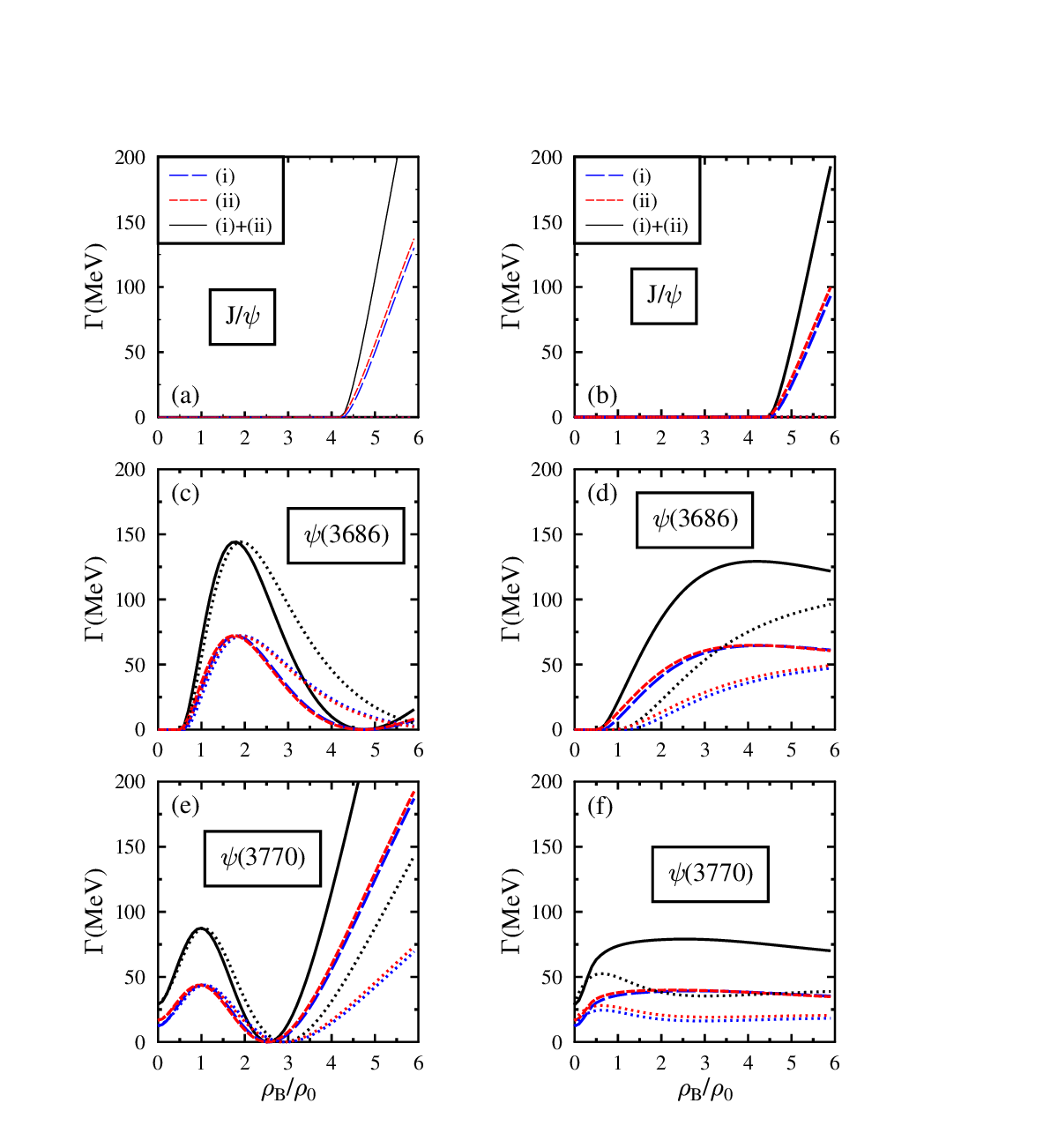}
} 
\caption{(Color online) The partial decay widths of the charmonium states, 
calculated using the $^3 P _0$ model, 
to (i) $D^+D^-$, (ii) $D^0\bar {\rm D^0}$ and (iii) the sum of the 
two channels ((i)+(ii)) in the isospin symmetric strange hadronic matter 
($\eta$=0,$f_s$=0.5), 
accounting for the medium modifications of the $D(\bar D)$ mesons. 
These are shown in subplots (a), (c) and (e), when the mass modifications 
of the charmonium states are neglected and (b), (d) and (f), 
the partial decay widths are shown when the mass modifications 
of the charmonium states are also taken into account. These results
are compared to the case of $f_s$=0, shown as dotted lines.}
\label{charmdw3p0eta0}
\end{figure}

\begin{figure}
\vskip -0.5in
\resizebox{0.8\textwidth}{!}{%
  \includegraphics{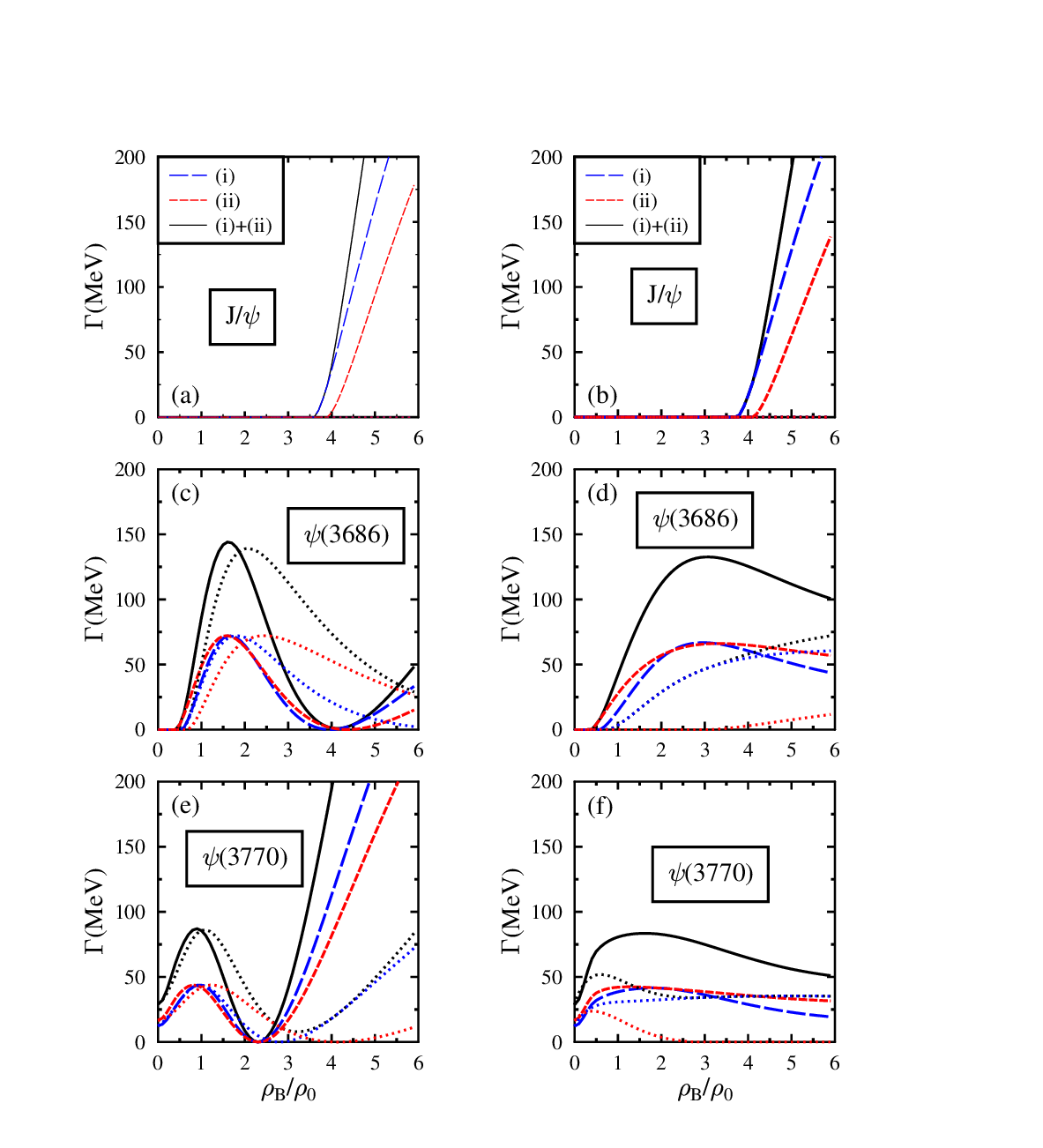}
} 
\caption{(Color online) The partial decay widths of the charmonium states, 
calculated using the $^3 P _0$ model, 
to (i) $D^+D^-$, (ii) $D^0\bar {\rm D^0}$ and (iii) the sum of the 
two channels ((i)+(ii)) in the isospin asymmetric strange hadronic matter 
($\eta$=0.5, $f_s$=0.5), as functions of the baryon density,
in units of the nuclear matter saturation density, 
accounting for the medium modifications of the $D(\bar D)$ mesons. 
These are shown in subplots (a), (c) and (e), when the mass modifications 
of the charmonium states are neglected and (b), (d) and (f), 
the partial decay widths are shown when the mass modifications 
of the charmonium states are also taken into account. These results
are compared to the case of $f_s$=0, shown as dotted lines.}
\label{charmdw3p0eta5}
\end{figure}

\begin{figure}
\resizebox{0.8\textwidth}{!}{%
  \includegraphics{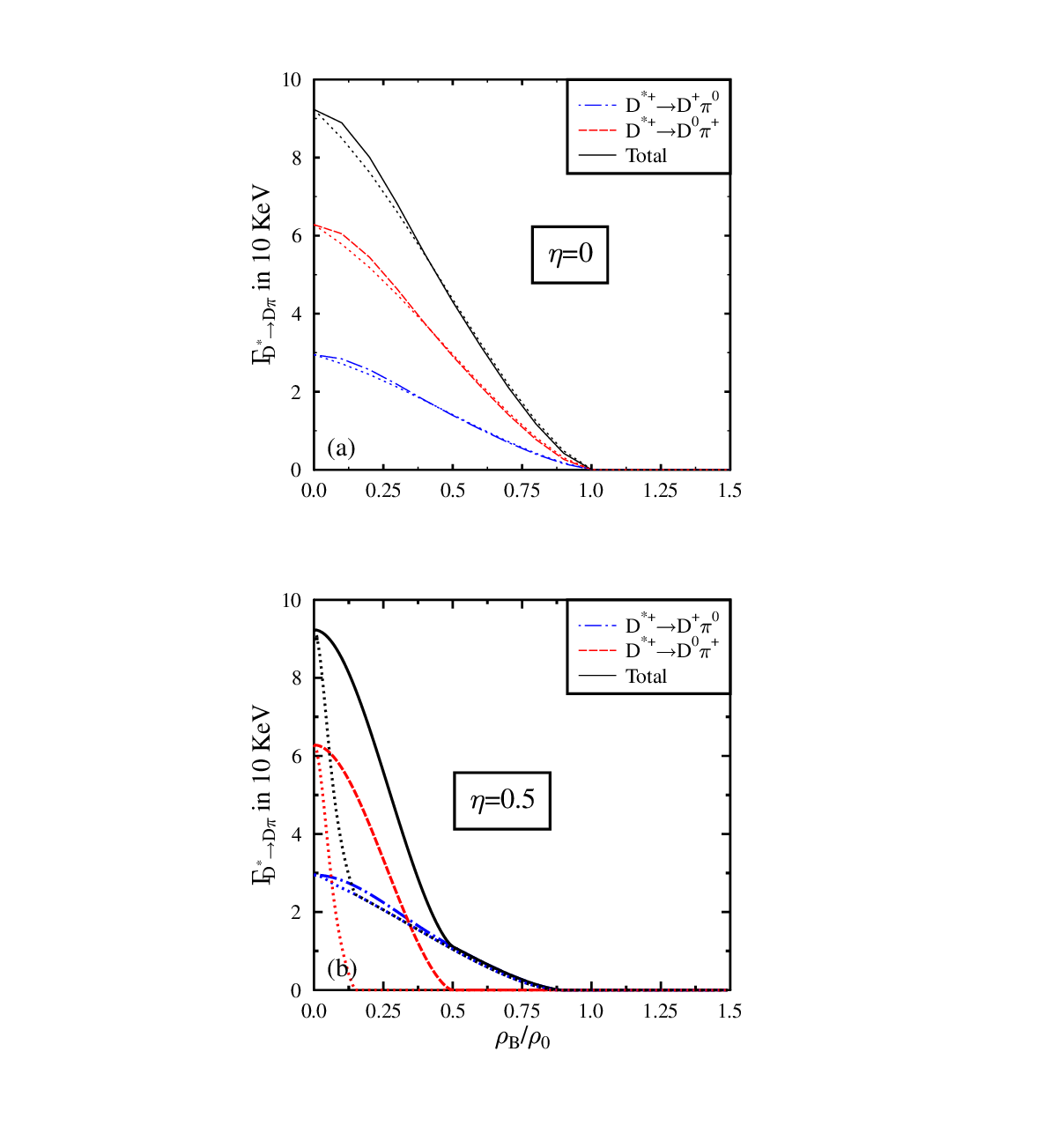}
} 
\caption{(Color online) 
The decay width $D^{*+}\rightarrow D\pi$, arising from the
decay processes, $D^{*+}\rightarrow D^+\pi^0$,
$D^{*+}\rightarrow D^0\pi^+$, plotted as a function of
the baryon density in units of nuclear matter saturation density
for the isospin symmetric as well as isospin asymmetric hyperonic
matter in subplots (a) and (b) respectively. The results are compared
with the case of nuclear matter.
}
\label{dstrdpi}
\end{figure}

\section{Results and Discussions}
In this section, we discuss the results of the modifications of the 
decay widths of the charmonium states, $J/\psi$, $\psi(3686)$ and 
$\psi(3770)$ to $D\bar D$ pair in isospin asymmetric strange hadronic matter,
using the field theoretic model for the composite hadrons described in section II. 
The medium modifications of the masses of $D$, $\bar D$ mesons and of the 
charmonium states calculated in a chiral effective model as in
section III have been discussed in detail in Ref. \cite{amcharmdw}.
These masses determine the in-medium decay widths.
We therefore plot these in figures \ref{mdasym}
and \ref{dmcharmasym}, and use them to calculate the decay widths as given 
in subsequent figures \ref{charmdecayeta0} and \ref{charmdecayeta5}.

In the field theoretic model here the decay widths are through
a pair creation Hamiltonian that arises naturally in the Dirac Hamiltonian
of the constituent field operators as in equation (\ref{hint}) from
Refs. \cite{spm781,spm782} with Lorentz boosting parallel to a translation 
as in Ref. \cite{spmdiffscat}.
The constituent masses of the light quarks (u,d), as has already been
mentioned earlier, are taken as 
${\rm {M_u}}={\rm {M_d}}$=330 MeV.
The constituent quark mass of the heavy charm quark is taken as  
$\rm {M_c}$=1600 MeV. The parameter $\lambda_2$, which is the energy fraction 
of the heavy quark (antiquark) in the $D (\bar D)$ mesons, is calculated
as in equations (\ref{omega2}) and (\ref{lambda12}), and is estimated as 
0.85.
In the present work, the values of the $R_{J/\psi}$, $R_\psi'$ 
and $R_{\psi''}$ of the strengths of the harmonic oscillator wave 
functions for the charmonium states 
are fitted so that the  root mean squared radii, $r_{rms}^2$ of the
charmonium states, $J/\psi$, $\psi'\equiv \psi(3686)$
and $\psi'' \equiv \psi(3770)$ are given as (0.47 fm)$^2$,
(0.96 fm)$^2$ and (1 fm)$^2$ respectively
\cite{leeko,amcharmdw}.
These values are obtained as $R_{J/\psi}$=(520 MeV)$^{-1}$, 
$R_{\psi'}$=(390 MeV)$^{-1}$, and $R_{\psi''}$
=(370 MeV)$^{-1}$. The values of $R_D$ and $\gamma$  
are taken as (310 MeV)$^{-1}$ and 0.35 respectively, so that
the decay widths of the $\psi '' \rightarrow D\bar D$ 
and partial decay widths of $\psi (4040)$ to $D\bar D$,
$D {\bar D}^*$, $D^* \bar D$ and $D^* {\bar D}^*$ 
in vacuum are reproduced in the $^3P_0$ model \cite{leeko,amcharmdw}. 
The above values of the  parameters $R_{J/\psi}$, $R_{\psi'}$,
$R_{\psi''}$, $R_D$ and $\gamma$ are used for the study
of the in-medium decay widths of the charmonium states
to $D\bar D$ using the $^3P_0$ model \cite{amcharmdw}.
With the above values of the harmonic oscillator strength
parameters for the charmonium states and $D(\bar D)$ mesons, 
the constituent quark masses for the light (u,d) quark and c-quark, 
and the value of the coupling strength, $\gamma_\psi$ as 1.35,
in the present field theoretic model for composite hadrons,
one reproduces the vacuum decay widths for the decay channels
$\psi'' \rightarrow D^+ D^-$ and $\psi'' \rightarrow D^0 \bar {D^0}$
\cite{amcharmdw}, as 12 MeV and 16 MeV respectively \cite{pdg}.

In the present model, the in-medium decay widths for the charmonium state
$\psi$ decaying to $D\bar D$ pair are given by equations 
(\ref{gammapsidpdm}) and (\ref{gammapsid0d0b}),
their medium dependence being through the
magnitude of $\bf p$, the 3-momentum of $D(\bar D)$ mesons,
given in terms of the masses of the charmonium, the $D$ and 
the $\bar D$ mesons, as in equation (\ref{pd}). The expressions 
for these decay widths are given as a polynomial part multiplied 
by a gaussian contribution. The density dependence of the decay widths of 
the charmonium states
($J/\psi$, $\psi'\equiv \psi(3686)$ and $\psi'' \equiv\psi(3770)$) 
to $D\bar D$ for isospin symmetric ($\eta$=0) hyperon matter 
with $f_s$=0.5 
are shown in figure \ref{charmdecayeta0} and compared to the case 
of symmetric nuclear matter ($\eta$=0,$f_s$=0). 
In isospin symmetric matter, the masses of the $D^0$ and $D^+$ 
of the $D$ meson doublet, as well as the masses of the $\bar {D^0}$
and $D^-$ of the $\bar D$ doublet, remain almost degenerate 
with the small mass difference between them through solution of their
dispersion relations arising due to the small mass difference 
in their vacuum masses. Hence, the partial decay widths of the 
charmonium state decaying to $D^+D^-$ and $D^0\bar {D^0}$ are almost
identical in isospin symmetric matter, as 
in figure \ref{charmdecayeta0}. For hyperonic matter,
the decay channel of $J/\psi$ to $D\bar D$ ($D^+D^-$ or $D^0\bar {D^0}$) 
becomes possible at densities higher than 4.2 $\rho_0$ when the in-medium
masses of $D$ and $\bar D$ mesons are taken into account, but the mass
of $J/\psi$ is taken to be its vacuum value. This value of the threshold 
density is modified to 4.5$\rho_0$, when the mass shift
of $J/\psi$ is also included. This is due to the fact that $J/\psi$ 
has only a small modification in the medium, as can be seen 
from the figure \ref{dmcharmasym}. For $J/\psi$, there is seen to be
an increase of the polynomial part of the decay width with density, 
which dominates over the gaussian part, thus leading to a monotonic rise 
in the decay of $J/\psi$, as seen in the subplots (a) and (b)
of figure \ref{charmdecayeta0}. However, these decay widths 
are observed to be very small in magnitude, being of the order of 
around 0.3 MeV at a density as large as 5$\rho_0$. 

The decay widths of $\psi(3686)$ and $\psi(3770)$ to $D\bar D$ 
are illustrated in subplots (c) and (e) of figure \ref{charmdecayeta0} 
for isospin symmetric hadronic matter ($\eta$=0), 
with the medium modifications of the masses of the $D$ and $\bar D$ mesons,
but not of the charmonium state. Due to drop in the mass
of $D\bar D$ pair in the medium, the decay of $\psi(3686)$ to $D\bar D$ 
becomes kinematically admissible above a density of around 0.6$\rho_0$,
as seen in subplot (c). The decay of $\psi(3770)$ to $D\bar D$
is already possible in vacuum, which however is modified in the medium
due to mass drop of the $D\bar D$ pair. As has already been mentioned,
the medium modifications of the decay widths are through the magnitude 
of  $\bf p$, the 3-momentum of the $D(\bar D)$ meson. For both of the 
excited charmonium states, $|\bf p|$ is seen to increase with density,
when only the mass modifications of the $D$ and $\bar D$ mesons are considered. 
This leads to an increase in the polynomial part of the decay width upto 
a density of about 2.3$\rho_0$ (1.6$\rho_0$) for $\psi(3686)$ ($\psi(3770)$), 
when the decay widths of $\psi(3686)$ and $\psi(3770)$ attain values 
of around 34 MeV and 169 MeV respectively.
As the density is further increased, the gaussian parts dominate 
due to increase in $|\bf p|$ with density, thus leading to a drop 
in their decay widths. The fall off with density is observed to be slower 
for the nuclear matter ($f_s$=0) as compared to for hyperonic matter.
This is due to the fact that $|\bf p|$ has a higher value in hyperonic matter 
as compared to in nuclear matter, since the masses of the $D$ and $\bar D$ 
mesons have smaller in-medium masses with strangeness in the medium.
When the in-medium masses of the charmonium states 
are also included, the partial decay widths of the excited
charmonium states, $\psi(3686)$ and $\psi(3770)$ (plotted in subplots 
(d) and (f)), are seen to be modified significantly. For hyperonic matter,
there is seen to be initially a sharp rise in the decay width with density
(reaching a value of around 25 MeV at a density of 4$\rho_0$ for 
$\psi(3686)$, and around 100 MeV at a density of $\rho_0$ for $\psi(3770)$). 
The increase, however, is observed to be a slower as the density 
is further increased. This is due to the fact that $|\bf p|$ 
becomes smaller with the in-medium charmonium mass, which leads to much 
lesser suppression arising from the gaussian part of the decay width 
at high densities, as compared to the case when the charmonium mass 
modification is not taken into account. 

We then consider the effects of the isospin asymmetry on the charmonium
decay widths. The masses of the $D^+$ and $D^0$, as well as for $D^-$
and $\bar {D^0}$ are no longer degenerate in the asymmetric hadronic medium,
thus leading to the charmonium decays to $D^+D^-$ and $D^0 \bar {D^0}$ 
to be different. In figure \ref{charmdecayeta5}, the decay widths of 
the charmonium states are shown for asymmetric hadronic matter with $\eta$=0.5.
These are plotted for the hyperonic matter with strangeness fraction $f_s$=0.5
and compared to the values for nuclear matter ($f_s$=0). When the medium 
modifications of the $D$ and $\bar D$ mesons are considered, but not 
of $J/\psi$, then the decay channels to (i) $D^+D^-$ and (ii)
$D^0 \bar {D^0}$ are observed to be admissible at densities higher than
around 3.6$\rho_0$ and 4$\rho_0$ respectively, as can be seen from subplot
(a) of figure \ref{charmdecayeta5}. The density dependence of the decay widths
still remains similar to the case for the isospin symmetric hadronic matter 
plotted in figure \ref{charmdecayeta0}. However, the values of the
decay widths turn out to be larger in magnitude as compared to 
case of the symmetric hadronic matter, with values of the order
of 0.5 MeV at a density of about 5$\rho_0$. There is seen to be
very small change in the decay widths when the in-medium mass for $J/\psi$
is included, with the threshold values for the decay to $D^+D^-$ 
and $D^0{\bar {D^0}}$ being modified to 3.7$\rho_0$ and 4.1$\rho_0$
respectively, as can be seen in subplot (b) of figure \ref{charmdecayeta5}.

The decay widths of the excited charmonium states $\psi(3686)$ and 
$\psi(3770)$ are plotted as functions of density in (c) and (e) of 
figure \ref{charmdecayeta5}, including the medium modifications of 
the masses of the D and $\bar D$ mesons, but not of the charmonium states. 
The density dependence of these decay widths are seen to be similar to
the isospin symmetric matter shown in figure \ref{charmdecayeta0}. 
In nuclear matter, the isospin asymmetry dependence is observed to be
quite appreciable leading to very different values for the charmonium 
decay widths to $D^+D^-$ and $D^0 \bar {D^0}$. This is due to the reason that
in the presence of hyperons, the mass modifications of the $D$
and $\bar D$ masses are lessened, as can be seen from figure \ref{mdasym}.
This leads to a much lesser effect from the isospin asymmetry in the hyperonic
medium as compared to in nuclear matter, as can be seen from the figure.
With in-medium charmonium masses, the value of $|\bf p|$ becomes smaller, 
leading to slower drop of the decay width of the charmonium states 
at higher densities, as can be seen from the
subplots (d) and (f) of figure \ref{charmdecayeta5}.
In asymmetric nuclear matter ($\eta$=0.5), 
the charmonium state $\psi(3686)$ is observed to decay to $D^+D^-$ 
at densities higher than about $\rho_0$, whereas the decay 
to $D^0 \bar {D^0}$ becomes possible at densities higher than
about 3.6$\rho_0$. The decay of $\psi(3770)$ to $D^+D^-$ 
is seen to have an initial rise upto a density of about 2$\rho_0$
reaching a value of about 38.5 MeV at 2$\rho_0$,
followed by a slow increase as the density 
is further increased. On the other hand, the decay to $D^0 \bar {D^0}$ 
does not become admissible above a density of about 2.5 $\rho_0$.
In hyperonic matter, the density behaviour of the decays of
$\psi(3686)$ and $\psi(3770)$ remain similar to the case of
symmetric hadronic matter plotted in figure \ref{charmdecayeta0}
and with inclusion of in-medium charmonium masses, the isospin
dependence still remains small for the hyperonic matter.

In the above, we have studied the in-medium decay width 
of the charmonium to $D\bar D$ pair, using the strength parameters
of the wave functions of the charmonium states to be fitted to
their root mean squared radii as (0.47 fm)$^2$, (0.96 fm)$^2$
and 1 fm$^2$ for $J/\psi$, $\psi'$ and $\psi''$ respectively. 
Let us investigate the sensitivity of the decay width 
with the size parameter of the charmonium state. 
In symmetric nuclear matter ($\eta$=0, $f_s=0$), 
it is observed that, when the value of $R_{\psi'}$ is increased 
(decreased) by 10\%, 
the decay width for $\psi'
\rightarrow D\bar D$ (in MeV) is modified from 
1.31$\times 10^{-2}$ 
to  1.46$\times 10^{-2}$
(9.65$\times 10^{-3}$) 
at density of $\rho_0$, 
and from 2.34 to 2.65 
(1.73)
at 2$\rho_0$ and from 
10 to 11.25
(7.63) 
at a density of 4$\rho_0$.
Hence one observes a modification of the decay
width of $\psi' \rightarrow D\bar D$ to be about
11-13\% larger, when the parameter $R_{\psi'}$ is increased by
10\% and about 23-25\% drop when $R_{\psi'}$ is decreased
by 10\%. The modification of the decay width of $\psi'$
remains similar in the presence of isospin asymmetry as well as
the strangeness in the medium. The modification of the 
decay width of $\psi'' \rightarrow D\bar D$ is also 
observed to be similar when $R_{\psi''}$ changes by about 
10\%. 
%

The partial decay widths of the charmonium states as calculated 
in the present composite hadronic model are now compared with the earlier 
results using the $^3 P_0$ model \cite{barnes,friman,amcharmdw}. 
For this purpose we show the decay widths obtained  from the $^3P_0$ model
\cite{amcharmdw} in figures \ref{charmdw3p0eta0} and \ref{charmdw3p0eta5} 
for the isospin symmetric ($\eta$=0) and isospin asymmetric ($\eta$=0.5) 
strange hadronic matter for $f_s$=0.5. As earlier,
the medium modifications of the decay widths are through $|\bf p|$, 
which is given in terms of the masses of the charmonium state, 
$D$ and $\bar D$ mesons by equation (\ref{pd}). The expressions 
for the decay widths of the charmonium states in the $^3 P_0$ model 
are also of the form of a polynomial part multiplied by a gaussian part
\cite{friman,barnes,amcharmdw}.
The density dependence of the charmonium decay 
width is thus due to the combined effect of these contributions.
For the excited charmonium states, $\psi(3686)$ and $\psi(3770)$, 
when the medium modifications of the $D$ and $\bar D$ mesons are considered,
but not of the charmonium state, then the charmonium decay widths
are initially seen to increase with density and then drop with further
increase in the density. This behaviour on density leads to even 
vanishing of the decay widths at certain densities \cite{amcharmdw}. 
There is seen to be an increase again at still higher densities.
For both symmetric as well as asymmetric matter, the decay widths 
of $\psi(3686)$ as well as $\psi(3770)$ in the $^3P_0$ model, 
are seen to vanish (see (c) and (e) of figures \ref{charmdw3p0eta0}
and \ref{charmdw3p0eta5}). 
Such nodes in the decay widths arising due to mass drop of the $D$ and 
$\bar D$ mesons in the medium, have been discussed earlier in the literature
\cite{friman} for excited charmonium states decaying to $D\bar D$ pairs.
A similar trend of the decay width with density
is still observed when the mass modifications
of the charmonium states are considered. 
However, no nodes are any longer observed 
even upto a density of about 6$\rho_0$.
The decay width of $J/\psi$ becomes kinematically accessible at densities
higher than about 4$\rho_0$ (4.5$\rho_0$) with the in-medium masses of
the $D$ and $\bar D$ mesons, and 
without (with) the mass modification
of $J/\psi$. In isospin asymmetric hadronic matter,
the decay widths for the charmonium state to $D^+D^-$ is observed 
to be different from the decay to $D^0 \bar {D^0}$ pair, 
due to difference in the masses of $D^+$ and $D^0$ 
of the $D$ meson doublet and of $D^-$ and $\bar {D^0}$
of the $\bar D$ doublet in the medium. Similar to the charmonium
decay widths of the present model for asymmetric matter 
shown in figure \ref{charmdecayeta5}, the $^3P_0$ also has
a much stronger dependence on the isospin asymmetry in nuclear
matter as compared to in hyperonic matter. This is due to the 
fact that the isospin asymmetry effect decreases with strangeness
in the hadronic medium. 

The qualitative behaviour of the partial decay widths of the 
charmonium states to $D\bar D$ with density, calculated within the present
composite hadron model, remains similar to their density dependence
as calculated using the $^3 P_0$ model.
The values obtained for the decay width of $\psi(3770)$ calculated
in the present investigation are very similar to those obtained 
by using the $^3P_0$ model.
However, the values of the decay widths of the charmonium
states, $\psi(3686)$ and $J/\psi$, are observed to be much smaller 
in magnitude as compared to those obtained in the $^3 P_0$ model.
Within the $^3P_0$ model, there are seen to be nodes in the decay widths
of the charmonium states, $\psi(3686)$ as well as $\psi(3770)$
when the in-medium masses of the $D(\bar D)$ mesons are considered, 
but the medium modification  of the charmonium mass is neglected
\cite{friman,amarvind}. However, in the present work, the decay
widths do show a behaviour of an initial increase with density and
then a drop, but no nodes are observed even upto a density of around
6$\rho_0$.

The decay width of $D^* \rightarrow D\pi$ in the hadronic medium 
has also been calculated in the present composite model for the hadrons.
To calculate this decay width, the value of the harmonic oscillator 
strength for $D^*$ meson, $R_{D^*}$ has been taken to be the same as
that of the $D$ meson, $R_D$ and the value of $R_{\pi}$ 
of the pion wave function is fitted to the square of the
charge radius of pion as $0.4 {\rm fm}^2$, which yields
$R_\pi$=(211 MeV)$^{-1}$ \cite{spm782}. 
The production strength, $\gamma_{D^*}$ for the creation of
$D\pi$ from $D^*$ is fitted from the decay widths,
$D^{*+} \rightarrow D^+ \pi^0$ and $D^{*+}\rightarrow D^0 \pi^+$
of 29.5 keV and 65 KeV respectively, and its value is estimated
to be 4.3. 
It may be noted here that the parameter
$\gamma_{D^*}$ corresponding to the decay $D^* \rightarrow D\pi$  
is higher than the value of the parameter $\gamma_\psi$ 
corresponding to the decay $\psi \rightarrow D\bar D$.
It was observed within the $^3P_0$ model \cite{barnes}
that with the value of the harmonic oscillator strength
$\beta \simeq$ 0.4 GeV, the parameter $\gamma$ 
turns out to be of the order of 0.5 for the decay
of the 1S and 1P light mesons, and of the order 
of 0.4 for higher L excited light mesons. 
In the presence of heavy quarks, e.g., c-quark,
e.g. for the decay of charmonium state to $D\bar D$ pair,
the value of $\gamma$ is observed to decrease 
so as to fit to the observed decay width
of $\psi (3770) \rightarrow D\bar D$ 
\cite{friman,leeko}. In Ref. \cite{friman},
where the strengths of the harmonic
oscillator wave function of the $D$-meson and the
charmonium state are chosen as $R_D$=(310 MeV)$^{-1}$
and $R_\psi=R_D/1.04$, for $\psi \equiv J/\psi,\psi',\psi''$),
the value of $\gamma$ turns out to be 0.281.
In Ref. \cite{leeko}, with $R_{\psi''}$ as fitted from its root mean
squared radius, $r_{rms}^2$=1 fm$^2$ and $R_D$=(310 MeV)$^{-1}$,
the value of $\gamma$ turns out to be 0.35. 
The general observation consistent with the
observed vacuum decay widths, within the $^3P_0$ model,
is that the pair production coupling constant
$\gamma$ decreases for the light meson sector,
when the decaying meson is with higher L state,
as well as for the case when the decaying meson
contains heavy quark (antiquark) as compared
to light quark (antiquark). 
In the present field theoretic model of composite hadrons
with quark constituents,
the decay of $D^*$ (heavy-light meson) is observed to
have a larger value for the pair production coupling
($\gamma_{D^*}$=4.3), as compared to for the decay 
of charmonium state ($\gamma_\psi$=1.35), 
which is a heavy-heavy meson. 
As has already been mentioned, we have adopted in the present work,
the harmonic oscillator strength parameters, $R_{J/\psi}$, $R_{\psi'}$, 
$R_{\psi''}$, to be fitted to their rms radii to be
(0.47 fm)$^2$, (0.96 fm)$^2$ and 1 fm$^2$ respectively, 
which yield their values as (520 MeV)$^{-1}$, (390 MeV)$^{-1}$
and (370 MeV)$^{-1}$, and taken the harmonic oscillator strength  
parameter for $D$ meson as $R_D$=(310 MeV)$^{-1}$ \cite{leeko}. 

The in-medium decay width has been plotted in figure 
\ref{dstrdpi} for the hyperonic matter (with $f_s$=0.5) for 
the isospin symmetric and asymmetric matter (with $\eta$=0.5).
The results have been compared with the situation of nuclear matter.  
For isospin asymmetric matter, the partial decay widths for
$D^{*+}\rightarrow D^+ \pi^0$ as well as  
$D^{*+}\rightarrow D^0 \pi^+$ are observed to decrease with 
increase in density and vanish above the nuclear matter saturation 
density. The effects from the hyperons is seen to be marginal 
for the isospin asymmetric hadronic matter. However, the effects
from strangeness fraction is seen to be appreciable for the case
of isospin asymmetric matter, as can been from the figure \ref{dstrdpi}.

\section{Summary and outlook}
We now very briefly summarise the results. 
We have here investigated the in-medium partial decay widths
of the charmonium states to $D\bar D$ pair, as well as
of the decay $D^* \rightarrow D\pi$ by using a composite
model for the hadrons with quark and/or antiquark constituents. 
The decay width is calculated by using quark pair creation term 
of the free Dirac Hamiltonian for the constituent quarks
within the composite hadron model. The matrix element
is multiplied with a strength parameter for the quark
pair creation, which is fitted from the observed vacuum
decay widths for the decays of $\psi''\rightarrow D\bar D$
and $D^* \rightarrow D\pi$.
The medium modifications of these partial decay widths 
are studied from the mass modifications of the charmonium and
$D(\bar D)$ mesons within a chiral effective model \cite{amcharmdw}.
When the mass modifications of the $D$ and $\bar D$ mesons are 
considered, but the charmonium mass taken as its vacuum value,
one finds for the excited charmonium states $\psi (3686)$ and 
$\psi(3770)$ that the decay width increases initially with density,
followed by a drop when the density is further increased.
There is seen to be an appreciable modification of these decay widths 
when the changes of the charmonium masses in the medium, are also 
taken into account. In the present work as well as in the 
$^3P_0$ model, the charmonium decay widths no longer seen to exhibit
any nodes (vanishing of these decay widths at certain densities),
when the medium modifications of the charmonium states are also
taken into account, in addition to taking the $D(\bar D)$ meson 
mass modifications into consideration. The isospin asymmetry 
effects are observed to be 
more dominant for the $D^+$ and $D^0$, as compared to $D^-$ and 
$\bar {D^0}$. This leads to the partial decay widths of charmonium states
to $D^+D^-$ and $D^0 \bar {D^0}$ to be different in asymmetric matter.
The isospin asymmetry effects are however observed to be quite dominant 
in nuclear matter and these effects are observed to be much less in 
the presence of hyperons in the medium.
The density effects on the decay widths of the charmonium states 
as well as $D^*$ mesons seem to be the dominant medium effects 
as compared to the effects of the isospin asymmetry and strangeness 
in the hadronic matter. These in-medium decay widths arise
from the mass modifications of the charmonium, $D$, $\bar D$ 
as well as $D^*$ mesons. 
The medium modifications of the masses of the charmonium states,
which are appreciable for the excited charmonium states, $\psi'$
and $\psi''$, as well as of the $D$, $\bar D$ and $D^*$ mesons
in the hadronic medium might be relevant in
in the compressed baryonic matter (CBM) experiments 
at the future facility of FAIR, GSI.
The interactions of the charmonium states and $D(D^*)$ 
with the nuclear medium could lead to the formation 
of exotic bound states of the nuclei with the (excited) 
charmonium states as well as with $D(D^*)$ mesons.
We note that parallel calculations could be useful for bottomonium
in the context of future experiments.

\begin{section}*{Acknowledgments}
The authors would like to thank Hiranmaya Mishra for discussions.
One of the authors (AM) is grateful to the Frankfurt Institute for 
Advanced Research (FIAS), University of Frankfurt, for warm hospitality 
and acknowledges financial support from Alexander von Humboldt Stiftung 
when this work was carried out. Financial support from Department of 
Science and Technology, Government of India (project no. SR/S2/HEP-031/2010)
is also gratefully acknowledged.
\end{section}


\begin{thebibliography}{}
\bibitem{kaplan1}
    D. B. Kaplan and A. E. Nelson, Phys. Lett. B {\bf 175}, 57 (1986).
\bibitem{kaplan2}
    A. E. Nelson and D. B. Kaplan, {\it ibid}, 192, 193 (1987).
\bibitem {blaiz}
	J. P. Blaizot and J. Y. Ollitrault, Phys. Rev. Lett.
	{\bf 77}, 1703 (1996).       	
\bibitem{satz}
	T. Matsui and H. Satz, Phys. Lett. B {\bf 178}, 416 (1986).  
\bibitem{haya1} Arata Hayashigaki , Phys. Lett. B {\bf 487}, 96 (2000).
\bibitem{qcdsum08}
T. Hilger, R. Thomas, B. K\"ampfer, Phys. Rev. C {\bf 79},
025202 (2009).
\bibitem{qmc1}
        K. Tsushima, D. H. Lu, A. W. Thomas, K. Saito, and R. H. Landau,
        Phys. Rev. C {\bf 59}, 2824 (1999).
\bibitem{qmc2}
        A. Sibirtsev,   K. Tsushima, and A. W. Thomas,
        Eur. Phys. J. A {\bf 6}, 351 (1999).
\bibitem{ltolos} L.Tolos, J. Schaffner-Bielich and A. Mishra,
Phys. Rev. {\bf C 70}, 025203 (2004).
\bibitem{ljhs}
L. Tolos, J. Schaffner-Bielich
and H. St\"ocker, Phys. Lett. {\bf B 635}, 85 (2006).
        \bibitem{mizutani6}
 T. Mizutani and A. Ramos, Phys. Rev. {\bf C 74},
065201 (2006).
\bibitem{mizutani8}
L.Tolos, A. Ramos and T. Mizutani, Phys. Rev. {\bf C 77},
015207 (2008).
\bibitem{HL}J. Hofmann and M.F.M.Lutz, Nucl. Phys. {\bf A 763}, 90 (2005).
\bibitem{sudoh} S. Yasui and K. Sudoh, Phys. Rev. {\bf C 87}, 015202
(2013).
\bibitem{dnkrein} C. E. Fontoura, G. Krein and V. E. Vizcarra,
Phys. Rev. {\bf C 87}, 025206 (2013).
\bibitem{hmamspm1} H. Mishra, S. P. Misra, Phys. Rev. D {\bf 48},
5376 (1993). 
\bibitem{hmamspm2} 
A. Mishra, S. P. Misra, Z. Phys. {\bf C 58} 325 (1993).
\bibitem{spmeffpot}A. Mishra, H. Mishra and S. P. Misra, 
Z. Phys. C {\bf 57}, 241 (1993).
\bibitem{spmbothcond}A.Mishra, H. Mishra, Varun Sheel, S. P. Misra,
P. K. Panda, Int. J. Mod. Phys. E {\bf 5}, 93 (1996). 
\bibitem{amhm2004} A. Mishra and H. Mishra, Phys. Rev. D {\bf 69},
014014 (2004).
\bibitem{kimlee}
Sugsik Kim, Su Houng Lee, Nucl. Phys. A {\bf 679}, 517 (2001).
\bibitem{charmmass2}
Arvind Kumar and Amruta Mishra, Phys. Rev. C {\bf 82}, 045207 (2010).
\bibitem{amcharmdw} Arvind Kumar and Amruta Mishra, Eur. Phys. Jour. 
{\bf A 47}, 164 (2011).
\bibitem{spm781} S. P. Misra, Phys. Rev. {\bf D 18}, 1661 (1978).
\bibitem{spm782} S. P. Misra, Phys. Rev. {\bf D 18}, 1673 (1978).
\bibitem{spmdiffscat} S. P. Misra and L. Maharana, Phys. Rev. 
{\bf D 18}, 4103 (1978).
\bibitem{friman} Bengt Friman, Su Houng Lee and Taesoo Song
 Phys. Lett. {\bf B 548}, 153 (2002).
\bibitem{yopr1}A. Le Yaouanc, L. Oliver, O. Pene and J.C. Raynal, 
Phys. Rev. {\bf D 8}, 2223 (1973). 
\bibitem{yopr2}A. Le Yaouanc, L. Oliver, O. Pene and J.C. Raynal, 
Phys. Rev. {\bf D 9}, 1415 (1974).
\bibitem{yopr3}A. Le Yaouanc, L. Oliver, O. Pene and J.C. Raynal, 
Phys. Rev. {\bf D 11}, 1272 (1975).
\bibitem{barnes} T. Barnes, F. E. Close, P. R. Page, E. S. Swanson,
Phys. Rev. {\bf D 55}, 4157 (1997).
\bibitem{spmddbar80} S.P.Misra, K. Biswal and B. K. Parida,
Phys. Rev. {\bf D 21}, 2029 (1980).
\bibitem{krein1}
        G. Krein, A. W. Thomas, K. Tsushima, Phys. Lett. B {\bf 697}, 
136 (2011).
\bibitem{paper3}
        P. Papazoglou, D. Zschiesche, S. Schramm, J. Schaffner-Bielich,
        H. St\"ocker, and W. Greiner, Phys. Rev. C {\bf 59},  411  (1999).
\bibitem{weinberg}
S.Weinberg, Phys. Rev. {\bf 166}, 1568 (1968).
\bibitem{coleman1}
S. Coleman, J. Wess, B. Zumino, Phys. Rev. {\bf 177} 2239 (1969).
\bibitem{coleman2}
C. G. Callan, S. Coleman, J. Wess, B. Zumino, Phys. Rev. {\bf 177}
2247 (1969).
\bibitem{bardeen}
W. A. Bardeen and B. W. Lee, Phys. Rev. {\bf 177}, 2389 (1969).
\bibitem{hartree}
        D. Zschiesche, A. Mishra, S. Schramm, H. St\"ocker and W. Greiner,
        Phys. Rev. C {\bf 70}, 045202 (2004).
\bibitem{kristof1}
        A. Mishra, K. Balazs, D. Zschiesche, S. Schramm,
        H. St\"ocker, and W. Greiner,
        Phys. Rev. C {\bf 69}, 024903 (2004).
\bibitem {amdmeson} 
A. Mishra, E. L. Bratkovskaya, J. Schaffner-Bielich, 
S.Schramm and H. St\"ocker, Phys. Rev. {\bf C 69}, 015202 (2004).
\bibitem{amarind} 
Amruta Mishra and Arindam Mazumdar, Phys. Rev. C {\bf 79},  024908 (2009). 
\bibitem{amarvind} 
Arvind Kumar and Amruta Mishra, Phys, Rev. C {\bf 81}, 065204 (2010).
\bibitem{sche1}
J. Schechter, Phys. Rev. D {\bf 21}, 3393 (1980).
\bibitem{cohen}
Thomas D. Cohen, R. J. Furnstahl and David K. Griegel,
 Phys. Rev. C {\bf 45}, 1881 (1992).
\bibitem{klingl} F. Klingl, S. Kim, S. H. Lee, P. Morath
and W. Weise, Phys. Rev. Lett. {\bf 82}, 3396 (1999).
\bibitem{pes1} M.E. Peskin, Nucl. Phys. {\bf B156}, 365 (1979).
\bibitem{leeko} Su Houng Lee and Che Ming Ko, Phys. Rev. C {\bf 67},
038202 (2003).
\bibitem{pdg}
J. Beringer et al (Particle Data Group), Phys. Rev. D {\bf 86},
010001 (2012).
\end{thebibliography}
\end{document}